# Signatures of magnetostriction and spin-phonon coupling in magnetoelectric hexagonal 15R-BaMnO$_3$

Bommareddy Poojitha, Anjali Rathore, Ankit Kumar, and Surajit Saha*

*Department of Physics, Indian Institute of Science Education and Research, Bhopal 462066, India*

*Correspondence: surajit@iiserb.ac.in*

**Abstract**

Spin-phonon coupling, the interaction of spins with surrounding lattice is a key parameter to understand the underlying physics of multiferroics and engineer their magnetization dynamics. Elementary excitations in multiferroic materials are strongly influenced by spin-phonon interaction, making Raman spectroscopy a unique tool to probe these coupling(s). Recently, it has been suggested that the dielectric and magnetic properties of 15R-type hexagonal BaMnO$_3$ are correlated through the spin-lattice coupling. Here, we report the observation of an extensive renormalization of the Raman spectrum of 15R-BaMnO$_3$ at 230 K, 280 K, and 330 K. Magnetic measurements reveal the presence of a long-range and a short-range magnetic ordering in 15R-BaMnO$_3$ at 230 K and 330 K, respectively. The Raman spectrum shows the appearance of new Raman modes in the magnetically ordered phases. Furthermore, an additional Raman phonon appears below ~ 280 K, possibly arising from a local lattice-distortion due to the displacement of Mn-ions, that exhibits anomalous shift with temperature. The origin of the observed renormalization and phonon anomalies in Raman spectra are discussed based on the evidences from temperature- and magnetic-field-dependent Raman spectra, temperature-dependent x-ray diffraction, magnetization, and specific heat measurements. Our results indicate the presence of magnetostriction and spin-phonon coupling in 15R-BaMnO$_3$ thus suggesting that the optical phonons are strongly correlated to its magnetoelectric properties.

**Introduction**

The coupling among various degrees of freedom such as spin, charge, lattice, and orbital is important as it gives rise to various novel phenomena and leads to exotic ground states in condensed matter systems [1-4]. The interplay between phonons and electron spins is relevant for many exotic phenomena including spin-peierls transition, phonon Hall effect, colossal magnetoresistance, ultrafast magnetization control, and spin Seebeck effect etc. [1-5] Recently, there has been a renewed interest in studying spin-phonon coupling (SPC) within the context of multiferroics and spintronics. In multiferroic materials, SPC has been implored to explain the compelling phenomena such as the thermal Hall-effect [6]. Moreover, SPC allows tailoring the functionalities of transition metal oxides such as stabilizing new multiferroic ground state by applying strain, tuning to unusual magnetic ground states etc. SPC also provides useful information about spin relaxation time which is an important concern in spintronics applications such as quantum computing [7-13]. Unfortunately, the number of candidate (multiferroic) materials is limited and it often happens that either the transitions are at very low temperatures or the response of cross-coupling between ferroic orders is very weak to be useful in device applications [14]. Hence, the search for new multiferroic materials with better functionalities is constantly increasing over the period of time.

In this context, the 15R-type hexagonal BaMnO$_3$ (15R-BaMnO$_3$) has received a renewed importance after the discovery of magnetoelectric property near room temperature [15], exhibiting antiferromagnetism below (T$_N$) ~ 230 K. It is considered to be an exceptional class of magnetoelectrics due to the fact that both the electric and magnetic properties are associated with the same Mn$^{+4}$ ions. The electric polarization arises from the displacement of the same ions (magnetic ions – Mn$^{4+}$) along the *c*-axis, *akin* to that of classic ferroelectric material BaTiO$_3$. It was believed that the asymmetric environment of Mn ions within MnO$_6$ Octahedra in the cubic and hexagonal layers leads to ferroelectric instability in the lattice which also has an intimate connection with magnetic ordering [15]. Korneta et al. observed that both magnetization and dielectric constant are highly anisotropic and the in-plane (ab-plane) dielectric constant is sensitive to an applied magnetic field [15]. They have attributed the observed correlation between magnetic and dielectric responses to the possible spin-phonon coupling in 15R-BaMnO$_3$. Since phonons play a major role in ferroelectric materials [16-18], one may expect them to have an equally important role in the magnetoelectric multiferroics too. A detailed understanding of the phonons and their coupling is, therefore, extremely important as it extends the opportunity to engineer new functionalities in transition metal oxides such as BaMnO$_3$ and others.

The idiosyncratic feature of spin-phonon coupling is the renormalization of phonon parameters. The electron spin can interact with its surrounding lattice in a number of possible ways, for instance, the lattice vibrations involving Mn-O or Mn-O-Mn bonds modify spin-spin correlations through superexchange interactions in antiferromagnets [19-21]. Similarly, if spins are coupled to elastic degrees of freedom, spontaneous strain occurs at magnetic transition temperature, giving rise to magnetostriction [22]. Multiferroic properties are associated with the lowering of symmetries (spatial and time-reversal symmetry). Since Raman spectroscopy is sensitive to these changes, it emerges as a powerful tool to detect the presence of coupling between different degrees of freedom, e.g., magnetostriction, spin-phonon coupling, and electron-phonon coupling etc. in strongly correlated multifunctional materials [11, 12, 22, 23]. In this article, we have investigated the correlation between the phonons and magnetoelectric properties of 15R-BaMnO$_3$. Temperature-dependent magnetization measurements reveal two magnetic transitions: a possible short-range ordering at T$_S$ ~ 330 ($\pm$ 0.2) K, hitherto unknown, and a long-range antiferromagnetic ordering at T$_N$ ~ 230 ($\pm$ 0.2) K. We have observed six Raman active modes responding to the magnetic transitions at T$_S$ (short-ranged ordering) and T$_N$ (antiferromagnetic ordering) which can be attributed to spin-phonon coupling and magnetostriction. Additionally, a new phonon mode emerges in the Raman spectra at temperatures below T$_D$ ~ 280 K, indicating a local lattice distortion possibly arising from displacements of the Mn-ions, the frequency which exhibits anomalous softening with decreasing temperature. Magnetic-field dependence of the Raman modes and temperature-dependent x-ray diffraction measurements corroborate our observations. The strength of spin-phonon coupling is estimated by a model based on mean-field and two-spin cluster approximations. The origin of a strong renormalization of the Raman active phonons below the magnetic transitions ($T_S/T_N$) is discussed in detail.

**Experimental details**

Polycrystalline samples of 15R-BaMnO$_3$ were synthesised by using solid-state reaction method. High purity BaCO$_3$ (99.999%) and MnO$_2$ (99.99%) (Sigma-Aldrich) powders were used as precursors. The stoichiometric mixture was ground well for 3 hours and calcined at 1250 °C and 1300 °C for 2 hours each with intermediate grindings. After the calcination, the resultant powder was ground, pelletized, and sintered at 1400 °C for 6 hours. Powder X-ray diffraction (PXRD) measurements were done by PANalytical Empyrean x-ray diffractometer with Cu-K$_\alpha$ radiation of wavelength 1.5406 Å. Liquid nitrogen-based Anton Paar TTK 450 heating stage was used to control the sample-temperature during non-ambient PXRD measurements. Chemical compositions were determined using energy dispersive x-ray (EDAX) technique equipped with high-resolution field emission scanning electron microscope (HR-FESEM) (Zeiss ULTRA Plus). Transmission electron microscopy (TEM) images were collected to verify the periodicity in the lattice using TALOS S-FEG (Schottky FEG (Field Emission Gun-Transmission Electron Microscope)) emitter with accelerating voltage of 200 kV. For the TEM measurement of the samples: a small amount of the powder sample of as synthesized compound was dispersed in distilled water by ultrasonicating for 20 s. This (turbid) solution was then drop-casted on a carbon coated copper grid forming a very thin layer. The solution/layer was then allowed to dry for a few hours thoroughly before it was attached to the sample holder on the microscope (TEM) for viewing. Raman spectra were collected from pelletized samples in the backscattering configuration using a LabRAM HR Evolution Raman spectrometer attached with a frequency-doubled Nd:YAG (neodymium-doped yttrium aluminium garnet; Nd:Y$_3$Al$_5$O$_{12}$) laser excitation source of wavelength 532 nm and Peltier cooled charge-coupled device (CCD) detector. A microscope objective of 50X magnification and numerical aperture of 0.5 was used to collect the scattered light. A Linkam stage (Model HFS600E-PB4) was used for variable temperature Raman measurements. The Raman spectrometer was also optically coupled to a closed cycle cryomagnet (Attocube 1000) to measure magnetic field-dependent Raman spectra at low temperature. DC magnetization measurements as a function of temperature and external magnetic field were carried out using Quantum Design SQUID-VSM (Superconducting Quantum Interference Device with Vibrating Sample Magnetometer). Specific heat ($C_p$) measurements were performed using a commercial Quantum Design PPMS on thin, flat pellet samples in the temperature range (T: 5 to 390 K). Data were collected in two different runs: the data from 5 to 295 K were collected in the first run where Apiezon N-grease was used (to stick the sample to the holder) as an addenda and the high temperature data (T: 295 to 390 K) were collected in the second run where it needs additional oven attachment and Apiezon H-grease (glue to stick the sample to holder) as an addenda. The data were collected during the heating cycle. Measurement on the sample was preceded by a measurement of the addenda in the same temperature range following a thermal relaxation method. Heat capacity ($C_p$) of the sample was obtained after subtracting the addenda heat capacity from total heat capacity.

**Results and Discussion**

*Structural and magnetic properties:*

The crystal structure of 15R-BaMnO$_3$, drawn using VESTA (Visualization of Electronic and STructural Analysis) software [24], is shown in Figure 1(a). It contains 15 formula units (75 atoms) per unit cell and 5 formula units (25 atoms) per primitive cell. 15R unit cell possesses an intermediate structure between an ideal cubic (c) and hexagonal (h) structures with a layer stacking sequence of (chhhh)$_3$ or in other words, atomic stacking of "ABCBACABACBAACB". The corner-shared connectivity (cubic to hexagonal stacking ratio) is 20%. It has Mn$_5$O$_{18}$ units (of five face-shared octahedra) connected to each other directly at their corners, forming a chain along the *c*-axis (Figure 1a). The crystal phase and corresponding unit cell parameters for the synthesized sample are obtained from Rietveld refinement of the room temperature x-ray diffraction patterns using High-score Plus software. The refinement indicates that the synthesized compound is in 15R type hexagonal phase possessing R-3m (No. 166) space group with rhombohedral symmetry. No additional or unexpected reflections could be identified thus confirming that the synthesized powder is of single-phase and free from any detectable impurity. The refined lattice parameters at room temperature are a = b = 5.6817 Å, c = 35.369 Å, α = β = 90°, and γ =120° which are comparable to the previous reports [15, 25, 26]. Three irreducible sites each for Ba, Mn, and O atoms in the unit cell which are labelled as Ba1, Ba2, Ba3, Mn1, Mn2, Mn3, and O1, O2, O3, respectively, are shown in Figure 1(a). The bond lengths and angles are listed in Table I. The refined crystallographic Wyckoff positions for each atom are listed in Table S1 in the supplemental material [27]. The Mn-Mn distance along the face-shared octahedra (2.4178 - 2.4783 Å) is comparable or smaller than the metallic γ-Mn (2.47 Å) which plays an important role in the physical properties of BaMnO$_3$ [28]. The strong electrostatic repulsion between Mn atoms leads to distortion in face-shared octahedra making them asymmetric while corner-shared octahedra remain symmetric with all the six Mn-O bonds of equal lengths (Table I) [15]. The spin arrangement in the magnetically ordered phase is displayed in Figure 1(b) which implies that the spins orient along the *a*-axis and are antiferromagnetically ordered along the *c*-axis [15] which will be discussed later. TEM images also confirm the formation of the lattice structure of the compound (Figure 1(c)). The *d*-spacing and lattice parameters are estimated and found to be in agreement with our PXRD results. Energy dispersive x-ray (EDAX) measurements were carried out for elemental analysis, details of which are given in Figure 1(d) and Table II. We found no detectable impurity elements or additional phases in the sample [see supplemental material for further details [27]].

To investigate the spin-ordering transitions, temperature-dependent magnetization measurements were carried out in two different protocols: zero-field cooling (ZFC) and field cooling (FC) under the applied field of 500 Oe as shown in Figure 2a. The transition temperatures can be clearly determined from the inflection in the derivative, $\frac{dM(T)}{dT}$, shown in the inset of Figure 2a. Our data show two magnetic transitions: one as an anomaly at 330 (± 0.2) K and another at 230 (± 0.2) K which are, respectively, the outcomes of a short-ranged direct exchange interaction (J$_D$) between Mn$^{+4}$ ions in face-shared MnO$_6$ octahedra of the Mn$_5$O$_{18}$ units and linear Mn-O-Mn superexchange (J1) between Mn$^{+4}$ ions located in corner-shared octahedra [29]. The bond angles for Mn2-O2-Mn3 and Mn3-O3-Mn1 are 81.07° and 78.28°, respectively, where the overlapping between 3d orbitals of $Mn^{+4}$ and O-2p orbitals is minimal, resulting in a relatively weak and negligible ferromagnetic (FM) exchange interaction which are denoted by J2 and J3, respectively (Figure 1b). Notably, the heat capacity as a function of temperature (Figure 2b) shows two anomalies (*kinks*) at ~ 330 (± 0.2) K and ~ 230 (± 0.2) K, corresponding to the two magnetic transitions thus corroborating our magnetization results. The observed low-temperature magnetic transition at ~ 43 (± 0.2) K is associated with spin canting (See Figures S1-S4 in section SM2 of Supplemental material [27] for further details including field-dependent magnetization results and further analysis on magnetic transitions) [30, 31]. Below 230 K ($T_N$), the spins orient along the *a*-axis and are antiferromagnetically ordered along the *c*-axis exhibiting an in-plane parallel arrangement with no direct exchange pathways, as shown in Figure 1b [spin alignment in full unit cell is shown in Figure S2 supplemental material [27]]. To understand the correlation between the phonons and magnetoelectric order parameters, we have performed a systematic Raman spectroscopic measurement with varying temperature and magnetic field as discussed below.

*Raman spectroscopy:*
According to group theory, the 15R structure with the space group R-3m leads to 18 Raman active and 27 IR active phonons at the Γ point of the Brillouin zone with irreducible representations $Γ_{Raman} = 8A_{1g} + 10E_g$ and $Γ_{IR} = 12A_{2u} + 15E_u$, respectively [Tables III & S2]. Because of the site symmetries, the atoms Ba1, Mn1, and O1 participate only in IR active vibrations and are not involved in Raman active modes, whereas, Ba2, Ba3, Mn2, Mn3, O2, and O3 take part in both IR and Raman active vibrations. Figure 3(a) shows the Raman spectra collected at room temperature (300K) and 80K along with their Lorentzian fits. The symmetry assignment of the modes is done based on our polarization-dependent Raman spectra (Figure S5 in supplemental material [27]), SYMMODES-Bilbao Crystallographic Server [32], and previous reports [26,33]. The spectrum at 80 K presents an additional number of modes as compared to the one collected at 300 K, implying the possible association with a change in the crystallographic and/or magnetic symmetry of $BaMnO_3$ at low temperatures [34-37]. To get a better understanding of the phonons, we have collected Raman spectra systematically in the temperature range of 80-800 K and analysed their evolution by fitting with Lorentzian multifunction. The Raman spectra at a few typical temperatures are shown in Figure S6 (Supplemental material [27]). The phonon parameters (Spectral weight/intensity, frequency, and linewidth) as a function of temperature for a few selected modes are displayed in Figure 3. The spectral weight (intensity) decreases with increasing temperature for almost all the modes, possibly because the temperature-dependence of intensity due to Bose factor is masked by the temperature-dependence of the absorption coefficient (intensity of the other modes is shown in Figure S7 in the supplemental material [27]) [38]. The frequency and linewidth of the majority of the phonons show the usual anharmonic trend (i.e. decrease (increase) in mode frequency (linewidth) with increasing temperature) as shown in Figure S8 and S9 in the supplemental material [27]. Notably, these modes exhibit a change in slope at one of the magnetic transition temperatures ($T_N$ or $T_S$) or at a temperature ($T_D$) ~ 280 K. More importantly, five of the phonon modes (namely, P7, P8, P12, P14, and P15) are observed to respond drastically at these temperatures as shown in Figure 3. The mode P7 at ~ 390 cm$^{-1}$ vanishes right above $T_D$ ~ 280

K. We attribute the origin of P7 mode to the displacement of $Mn^{+4}$ ions along the *c*-axis amounting to a change in local symmetry that may be associated with the dielectric anomaly reported in ref. [15]. The frequency (of mode P7) shows an anomalous temperature-dependence (i.e. a decrease in frequency with decreasing temperature) while the linewidth shows an extremely large broadening (by almost 4-fold) in the temperature range of 80-280 K. The anomalies in P7 could be related to spin-phonon coupling to be discussed later. On the other hand, the P12 and P14 modes appear right below the magnetic transitions $T_N$ and $T_S$, respectively. In addition, each of the modes at ~ 415 cm$^{-1}$ (P8) and ~ 653 cm$^{-1}$ (P15) clearly split into two modes right below $T_N$ ~ 230 (± 0.2) K. To note that our temperature-dependent x-ray diffraction measurements (to be discussed later) show no clear signatures of structural phase transition in the investigated temperature range (90 - 400 K) thus ruling out structural phase transition as the possible origin for the splitting of the two modes P8 and P15. However, possibilities of further changes in the local symmetries at $T_N$ arising from additional displacements of Mn ions cannot be completely ruled out. Further, the mode P6 deviates from the anharmonic trend (discussed below) at temperatures below $T_N$ (similar to the deviations seen for the other modes below $T_N/T_S/T_D$, shown in Figures S8 and S9 [27]). These observations suggest the presence of a strong correlation between phonons and magnetoelectric parameters in 15R-BaMnO$_3$. In order to understand the origin of phonon anomalies and appearance of new peaks in Raman spectra below magnetic transitions, we have analysed the temperature-dependence of phonon frequency and linewidth using the anharmonic model.

In general, the temperature-dependence of a phonon frequency can be written as [39-42]:

$$\omega(T) = \omega_{anh}(T) + \Delta\omega_{el-ph}(T) + \Delta\omega_{sp-ph}(T) \quad (1)$$

The first term $\omega_{anh}(T)$ is the anharmonic contribution as discussed below. The contribution to the change in phonon frequency due to electron-phonon coupling is given by the term $\Delta\omega_{el-ph}(T)$ which is absent in 15R-BaMnO$_3$ due to its insulating nature. The renormalization of the frequency due to spin-phonon coupling is accounted by $\Delta\omega_{sp-ph}(T)$. In the cubic anharmonic process, the phonon of frequency $\omega_0$ decays into two phonons of equal frequency $\omega_0/2$ satisfying the energy and momentum conservation. The temperature-dependence of the phonon frequency due to cubic anharmonicity (three-phonon process) is given by [39,40]:

$$\omega_{anh}(T) = \omega_0 + A\left[1 + \frac{2}{\left(e^{\frac{\hbar\omega_0}{2k_BT}} - 1\right)}\right] \quad (2)$$

Similarly, the temperature-dependence of phonon linewidth due to cubic anharmonicity can be written as:

$$\Gamma_{anh}(T) = \Gamma_0 + C\left[1 + \frac{2}{\left(e^{\frac{\hbar\omega_0}{2k_BT}} - 1\right)}\right] \quad (3)$$

where $\omega_0$ and $\Gamma_0$ are frequency and linewidth of the phonon at absolute zero temperature, *A*, and *C*, are cubic anharmonic coefficients for frequency and linewidth, respectively, $\hbar$ is reduced Planck constant, $k_B$ is Boltzmann constant and $T$ is the variable temperature. As seen in our data, both the frequency and the linewidth of almost all phonons show a finite deviation from the anharmonic temperature-dependence, indicating the possible magnetostriction in the system (Figure 3(c, d) & Figures S8, S9 [27]). The solid lines in Figure 3 (c, d) represent the fitting with Eq. 2 and 3. The mode P12 appears below $T_N$ (origin to be discussed later) and its frequency and linewidth exhibit the expected thermal behaviour (Eq. 2 and 3, respectively). Most importantly, the mode P7 that appears below $T_D$ shows an anomalous shift in frequency while its linewidth undergoes an anomalous broadening which cannot be explained by anharmonicity alone. These indicate a strong role of spin-phonon coupling. Further, the P6 and P14 modes show a deviation from anharmonic thermal behaviour in their temperature-dependence of frequency and linewidth at low temperatures below the Neel temperature $T_N \sim$ 230 K. In the absence of any structural phase transition, we attribute the observed anomalies in P6, P7, and P14 phonons to the spin-phonon coupling. The deviation in frequency ($\Delta\omega_{sp-ph}$) in modes P6, P7, and P14 from their anharmonic behaviours (see Figure 3(c)) with respect to temperature are captured in Figure 4. The strength of spin-phonon coupling is estimated using mean-field and two-spin cluster approximations as explained below.

*Spin-phonon coupling:*
In magnetic materials, when a phonon is coupled to the spin degrees of freedom, its frequency gets renormalized in proportion to the nearest-neighbour spin-spin correlation function $< S_i.S_j >$ (Schematically shown in Figure S10 in the supplemental material [27]) which can be written as [41-43]:

$$\Delta\omega_{sp-ph} = -\lambda_{sp} < S_i.S_j > = -\lambda_{sp}\Phi(T)S^2 \qquad (4)$$

where $\lambda_{sp}$ is the strength of spin-phonon coupling and $\Phi$ is the short-range order parameter. Lockwood et al. [42] theoretically estimated the $\Phi(T)$ for $S = 2$ ($FeF_2$) and $S = 5/2$ ($MnF_2$) antiferromagnetic systems using mean-field and two-spin cluster approximations. Since the estimates of $\Phi$ do not vary much with the value of the spin ($S$), it was also used reasonably for $S = 1$ ($NiF_2$ and NiO) antiferromagnets [41, 43]. Therefore, it is also reasonable to make use of the $\Phi$ estimated by Lockwood et al. [42] for $BaMnO_3$ in our case where $S = 3/2$ ($Mn^{4+}$). Thus, the spin-phonon coupling can be estimated using the relation [43]:

$$\lambda_{sp} = -\frac{\omega(T_{Low}) - \omega_{anh}(T_{Low})}{[\Phi(T_{Low}) - \Phi(2T_N)]S^2} \qquad (5)$$

where $\omega(T_{Low})$ is the experimental phonon frequency at the lowest temperature recorded ($T_{Low} \sim 80$ K in our case) while the $\omega_{anh}(T_{Low})$ is the corresponding anharmonic estimate of the phonon frequency at the same temperature. The obtained values for the spin-phonon coupling ($\lambda_{sp}$) for the modes P6, P7, and P14 using eqn. (5) are 1.2, 3.8, and 1.5 cm$^{-1}$, respectively. The coupling strength for the modes is reasonably strong and may be compared with some of the reported values of the strength in compounds like 9R-$BaMnO_3$ ($\lambda \sim 0.5$ to 3.4 cm$^{-1}$ [44]), 4H-$Sr_{0.6}Ba_{0.4}MnO_3$ ($\lambda \sim 2.2$ cm$^{-1}$ [45]), $MnF_2$ ($\lambda \sim 0.4$ cm$^{-1}$ [42]), $FeF_2$ ($\lambda \sim 1.3$ cm$^{-1}$ [42]), $La_2CoMnO_6$ ($\lambda \sim 1.7–2.1$ cm$^{-1}$ [46], $Pr_2CoMnO_6$ ($\lambda \sim 0.51–1.61$ cm$^{-1}$ [47,48]),

Cr$_2$Ge$_2$Te$_6$ ($\lambda \sim$ 0.24–1.2 cm$^{-1}$ [49]), Sr$_2$CoO$_4$ ($\lambda \sim$ 3.5 cm$^{-1}$ [50]), NiO ($\lambda \sim$ −7.9 cm$^{-1}$ and 14.1 cm$^{-1}$ for TO and LO phonons, respectively [43]), ZnCr$_2$O$_4$ ($\lambda \sim$ 3.2–6.2 cm$^{-1}$ [51]), NaOsO$_3$ ($\lambda \sim$ 40 cm$^{-1}$ [52]), and CuO ($\lambda \sim$ 50 cm$^{-1}$ [53]).

*Emergence of new Raman modes: Effect of magnetostriction*

Emergence of new modes in Raman spectra is possible below magnetic transition due to various reasons such as concurrent magnetic and structural phase transition [54], spin-excitations including one-magnon or two-magnon Raman processes [35, 55], or magnetostriction [36,37]. Since phonons and magnetic excitations exhibit contrasting behaviour under external magnetic fields, we have investigated magnetic-field-dependence of the Raman spectrum to shed light on the origin of the new modes. 15R-BaMnO$_3$ was cooled down to 4 K and the magnetic field was applied in the range of 0-9T. Figure 5a compares the Raman spectra at 0 T, 3 T, 6 T, and 9 T where no major effect on the spectrum is visible, thus we can rule out the magnetic excitation as the origin for the new modes and assign them as new phonon modes. Figure 5b shows the frequency of a few selected phonons as a function of the magnetic field (see also Figures S11-S14 in the supplemental material [27] for phonon parameters). The frequency of all the phonons shows a decreasing trend with increasing magnetic field thus suggesting a lattice expansion under an external magnetic field, indicating a possible role of magnetostriction.

In order to ensure any possible role of structural phase transition, on the origin of the new phonon modes, we have performed temperature-dependent powder x-ray diffraction measurements (PXRD). Figure 6a displays the PXRD patterns collected at a few temperatures in the range of 90-400 K. The number of reflections in the PXRD pattern remains the same in the entire investigated temperature range and Rietveld analysis of the data confirms the absence of structural phase transition.

In absence of a structural phase transition, the origin of the new phonon modes below the magnetic transition temperatures could be attributed to the local changes in the symmetry arising from the displacement of Mn-ions at low temperatures. The thermal expansion is a key parameter to ascertain the magnetostriction because it involves the coupling between magnetic and elastic degrees of freedom. We have analysed the unit cell parameters (lattice parameters and unit cell volume) as a function of temperature. Figure 6b shows the lattice parameters and unit cell volume as a function of temperature in the range of 90-450 K (bond lengths are given in Figure S15 in supplemental material [27]). The 15R-BaMnO$_3$ expands upon heating and shows positive thermal expansion. In general, the temperature-dependent lattice parameters due to thermal expansion can be written as [56]:

$$a(T) = a_0\left[1 + \frac{be^{\frac{d}{T}}}{T(e^{\frac{d}{T}}-1)^2}\right] \quad \text{and} \quad c(T) = c_0\left[1 + \frac{fe^{\frac{g}{T}}}{T(e^{\frac{g}{T}}-1)^2}\right] \quad (6)$$

where, $a_0$ and c$_0$ are the in-plane and out-of-plane lattice constants at 0 K, whereas b, d, f, and g are fitting parameters. Similarly, unit cell volume as a function of temperature can be expressed as:

$$V(T) = V_0 \left[1 + \frac{A}{(e^{\frac{\theta}{T}}-1)}\right] \quad (7)$$

where, $V_0$ is cell volume extrapolated to 0 K, $\theta$ is the Debye temperature, and $A$ is a fitting parameter (see Table S4 and Section SM7 in the supplemental material [27]). Importantly, the temperature-dependence of the lattice parameters and the volume show a deviation at $T_S$ as shown in Figure 6. As discussed earlier, a short-ranged magnetic ordering occurs below $T_S$ ~ 330 K that changes the trend of lattice expansion owing to local symmetry changes (without undergoing a structural phase transition) thus corroborating the presence of magnetostriction in 15R-BaMnO$_3$ [57].

It is to be noted that the frequency of a phonon $\omega \propto \sqrt{k}$ ($k$ = spring constant) where $k$ decreases with increasing unit cell volume (bond length). The decrease in phonon frequency with increasing magnetic field, as shown in Figure 5b, signify an expansion of the unit cell volume under magnetic field further suggesting the presence of magnetostriction in 15R-BaMnO$_3$. Here, we have quantified the magnetostriction in two ways: (1) Using thermal expansion (x-ray diffraction measurements) in the absence of external magnetic field and (2) using lattice expansion (Raman spectroscopic measurements) under external magnetic field at a constant temperature.

If the spontaneous distortion in the unit cell (without changing the crystal symmetry) due to magnetic ordering is large enough, it can be detected by x-ray diffraction technique which also provides several advantages over conventional strain-gauge and capacitive techniques [57-59]. The spontaneous volume magnetostriction ($\lambda_{ms}^V(T)$) at a given temperature '$T$' (below $T_N$) can be defined as [61]:

$$\lambda_{ms}^V(T) = \frac{V_{AFM}(T) - V_{PM}(T)}{V_{PM}(T)} \quad (8)$$

where $V_{AFM}(T)$ is the actual unit cell volume at the temperature $T$ in the antiferromagnetic phase whereas $V_{PM}(T)$ is the hypothetical volume of the unit cell if it were in the paramagnetic (nonmagnetic) phase at the same temperature. The value of $\lambda_{ms}^V$ at 90 K (lowest measured temperature) is ~ 32($\pm$ 7)x 10$^{-4}$, which is comparable to the reported value of magnetostriction in the spinel compounds Zn$_{1-x}$Cu$_x$Cr$_2$Se$_4$ (4.6 to 24.9x10$^{-4}$ at 100 K for varying x) measured by XRD technique [61] (temperature-dependent values of $\lambda_{ms}^V$ for 15R BaMnO$_3$ are given in Figure S16 in the supplemental material [27]).

On the other hand, magnetostriction can also be defined as the change in the dimensions of a magnetic material under an external magnetic field. It is often characterized by $\frac{\Delta l}{l}$ where $l$ is the original physical length of the material and $dl$ is the change in the dimension of the material under the applied magnetic field ($H$) [60]. It can be expressed as $\lambda_{ms}^l = \frac{\Delta l}{l} \approx \frac{l(H) - l(0)}{l(0)}$ for the linear magnetostriction and similarly, the volume magnetostriction is $\lambda_{ms}^V = \frac{\Delta V}{V} \approx \frac{V(H) - V(0)}{V(0)}$, where, for our case we have considered the unit cell dimensions instead of the physical dimensions. The change in unit cell volume can be related to the corresponding change in phonon frequency through Grüneisen parameter as $\frac{\Delta\omega}{\omega} = \gamma \frac{\Delta V}{V}$. In solids, the typical value of mode Grüneisen parameter ($\gamma$) is ~ 1 (see supplemental material [27] for details). Under the assumption of $\gamma \sim 1$, the volume magnetostriction of 15R-BaMnO$_3$ at 4 K is estimated to be in the range of 0.01 to 5.6x10$^{-4}$ (Figure S21 in supplementary material [27]), which is comparable to the range of reported magnetostriction values (10$^{-3}$ to 10$^{-6}$) in various systems [62-68] and also comparable to the value obtained through x-ray diffraction technique discussed above (see Figure S17-21 and Table S5 in the section SM8 of Supplemental material [27] for details on estimation of $\gamma$ and magnetostriction in 15R-BaMnO$_3$).

**Conclusions**

In summary, we have synthesized polycrystalline hexagonal BaMnO$_3$ in 15R phase using the solid-state reaction route. A new phonon of frequency ~ 390 cm$^{-1}$ (P7), associated with Mn vibrations, appears below ~ 280 K (due to local distortions arising from the displacement of Mn-ions) and exhibits anomalous behaviour due to a strong spin-phonon coupling ($\lambda_{sp\text{-}ph}$ ~ 3.8 cm$^{-1}$). Additionally, several new modes appear below the magnetic transition temperatures (330 K and 230 K) due to the local changes in symmetry arising from the displacement of Mn-ions. Further, we have observed evidences of magnetostriction in 15R-BaMnO$_3$ through a change in lattice expansion rate across the magnetic ordering temperature in XRD measurements and signatures of lattice expansion under applied magnetic field in Raman scattering. We have shown that the lattice vibrations are strongly correlated with the electric and magnetic properties in 15R-BaMnO$_3$. We believe that the spin-phonon coupling and lattice strain (magnetostriction) can be the new parameters to tune the magnetoelectric properties of the system thus making it potential for near room-temperature technological applications and motivating further experimental and theoretical studies.

**Acknowledgement**

Authors acknowledge IISER Bhopal for research facilities, B. P. acknowledges the University Grant Commission for fellowship and S. S. acknowledges DST/SERB (project No's ECR/2016/001376 and CRG/2019/002668) and Nanomission (Project No. SR/NM/NS-84/2016(C)) for research funding. Support from DST-FIST (Project No. SR/FST/PSI-195/2014(C) is also thankfully acknowledged. Authors acknowledge Mr. Sajilesh K P (Department of Physics, IISER Bhopal) for helping with the heat capacity measurements.

Table I. Lattice parameters, bond lengths, and bond angles of 15R-BaMnO$_3$ obtained from the Rietveld refinement of x-ray diffraction data at room temperature.

| Lattice parameters | a = b = 5.6817 Å, c = 35.369 Å, α = β = 90º, γ =120º | |
|---|---|---|
| Bond lengths | Mn2-Mn2 | 3.8406 Å |
| | Mn2-Mn3 | 2.4783 Å |
| | Mn3-Mn1 | 2.4178 Å |
| | Mn2-O1 | 1.9203 Å |
| | Mn2-O2 | 1.9294 Å |
| | Mn3-O2 | 1.8834 Å |
| | Mn3-O3 | 1.9248 Å |
| | Mn1-O3 | 1.9055 Å |
| Bond angles | Mn1-O1-Mn1 | 180º |
| | Mn2-O2-Mn3 | 81.07º |
| | Mn3-O3-Mn1 | 78.28º |

Table II. EDAX data of 15R-BaMnO$_3$.

| EDAX results | | |
|---|---|---|
| **Element** | **Weight%** | **Atomic%** |
| O K | 22.00(± 0.08) | 63.47(± 0.09) |
| Mn K | 20.43(± 0.13) | 17.17(± 0.13) |
| Ba L | 57.57(± 0.14) | 19.36(± 0.06) |
| Total | 100.00 | 100 |

Table III. Raman active phonons in 15R-BaMnO$_3$.

| Space group R-3m (No.166) and $\Gamma_{Raman} = 8A_{1g} + 10E_g$ | | Wyckoff positions | | Irreducible representations |
|---|---|---|---|---|
| | | 6c (Ba2, Ba3) | | $2A_{1g} + 2E_g$ |
| | | 6c (Mn2, Mn3) | | $2A_{1g} + 2E_g$ |
| | | 18h (O2, O3) | | $4A_{1g} + 6E_g$ |
| Mode | Symmetry | ω (cm$^{-1}$) | | Atoms involved in the vibration |
| | | at 80 K | at 300 K | |
| P1 | $E_g$ | 82.0 ($\pm$ 0.1) | 80.9 ($\pm$ 0.1) | Ba |
| P2 | $A_{1g}$ | 100.7 ($\pm$ 0.1) | 99.3 ($\pm$ 0.1) | Ba |
| P3 | $E_g$ | 176.5 ($\pm$ 0.1) | 174.7 ($\pm$ 0.1) | Ba |
| P4 | $E_g$ | 241.8 ($\pm$ 0.1) | 238.9 (0.1) | Mn |
| P5 | $E_g$ | 278.5 ($\pm$ 0.1) | 275.6 ($\pm$ 0.1) | Mn |
| P6 | $E_g$ | 361.4 ($\pm$ 0.1) | 358.4 ($\pm$ 0.1) | Mn |
| P7 | $A_{1g}$ | 383.5 ($\pm$ 0.2) (Appears at T < 280 K) | Absent | Mn |
| P8A | $A_{1g}$ | 418.2 ($\pm$ 0.1) | P8: 415.3 ($\pm$ 0.1) ($E_g$) | O |
| P8B | $A_{1g}$ | 422.0 ($\pm$ 0.1) (Appears at T < 230 K) | | O |
| P9 | $E_g$ | 430.3 ($\pm$ 0.1) | 427.5 ($\pm$ 0.1) | O |
| P10 | $A_{1g}$ | 535.6 ($\pm$ 0.4) | 530.0 ($\pm$ 0.4) | O |
| P11 | $E_g$ | 557.9 ($\pm$ 0.1) | 552.2 ($\pm$ 0.1) | O |
| P12 | $E_g$ | 565.9 ($\pm$ 0.1) (Appears at T <230 K) | Absent | O |
| P13 | $E_g$ | 583.0 ($\pm$ 0.1) | 578.8 ($\pm$0.1) | O |
| P14 | $A_{1g}$ | 643.2 ($\pm$ 0.1) | 636.9 ($\pm$0.3) | O |
| P15A | $A_{1g}$ | 655.8 ($\pm$ 0.1) | P15: 653.3 ($\pm$ 0.1) ($E_g$) | O |
| P15B | $A_{1g}$ | 659.9 ($\pm$ 0.4) (Appears at T < 230 K) | | O |

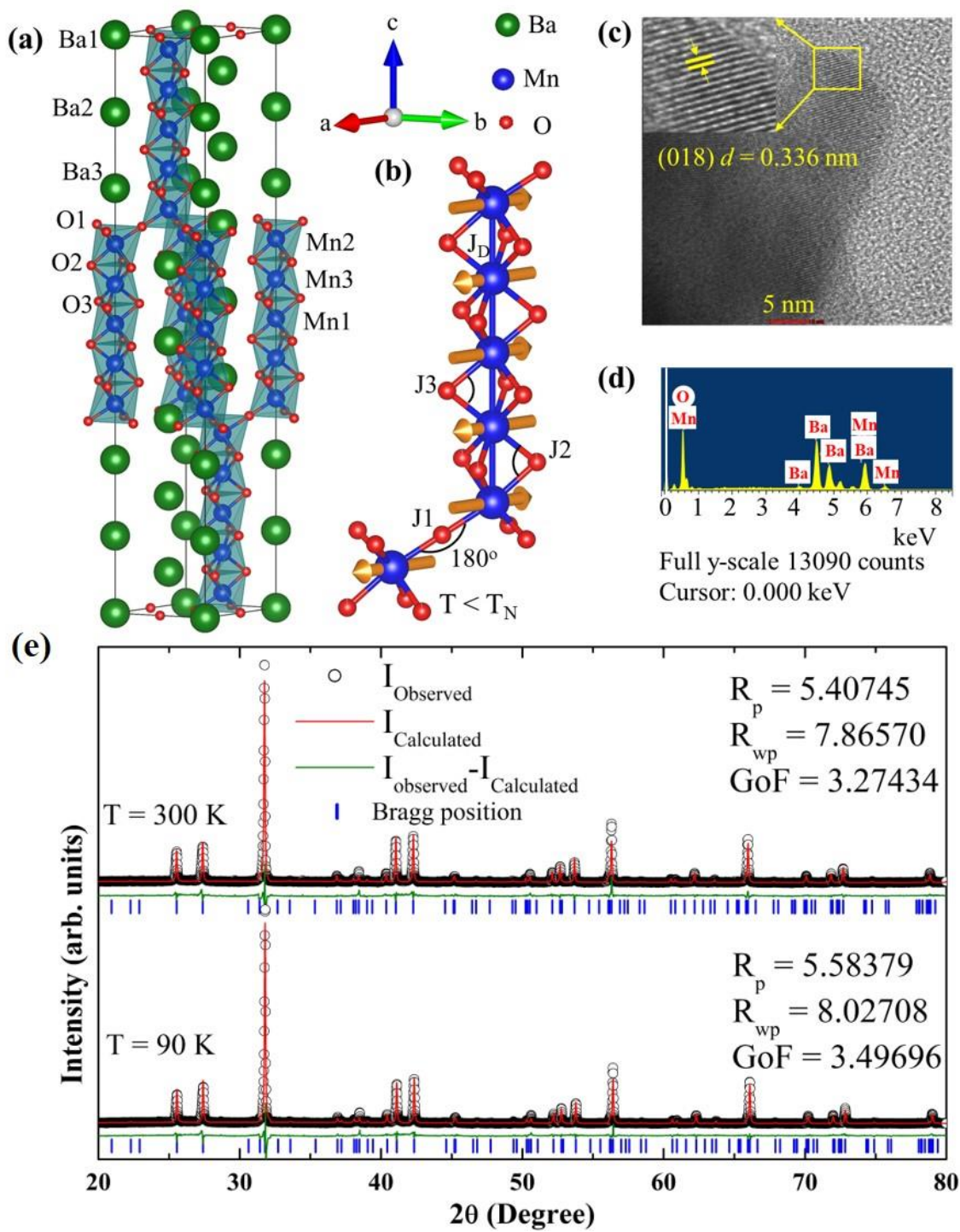

*Figure 1. (Colour online) (a) Crystallographic unit cell of 15R-BaMnO3, (b) spin arrangement of Mn-ions in antiferromagnetic phase, (c) TEM image showing the lattice planes, (d) EDAX spectrum showing the stoichiometric presence of the elements, and (e) X-ray diffraction patterns collected at 300 K and 90 K. R-factors ($R_p$ and $R_{wp}$) and the Goodness of Fit (GoF) obtained from the refinements are also given.*

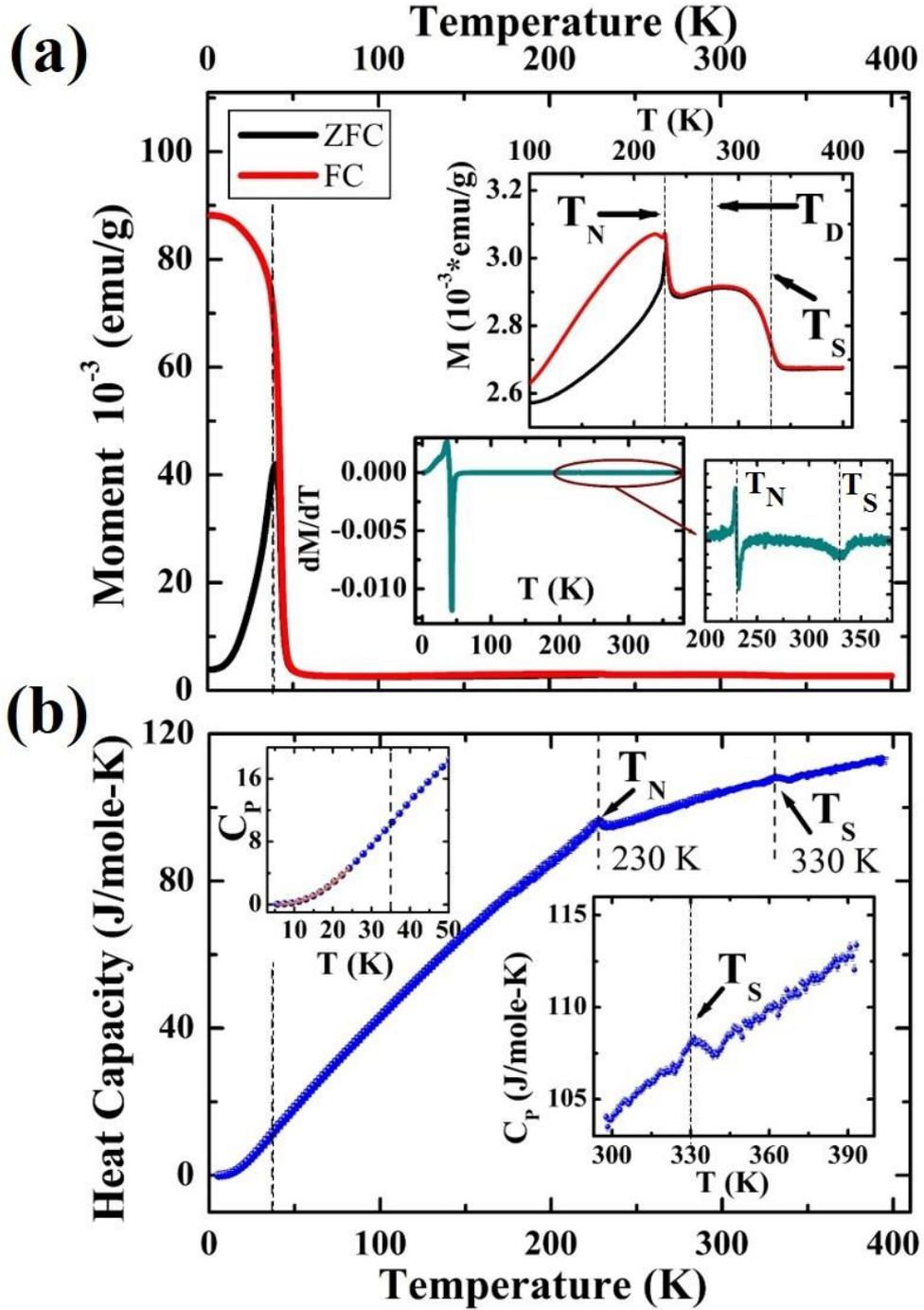

*Figure 2. (Colour online) (a) Magnetic moment as a function of temperature and insets show enlarged view in high temperature range and the temperature derivative of magnetization to clearly identify the antiferromagnetic transition temperatures, (b) temperature-dependent heat capacity, left inset is enlarged view of the heat capacity ($C_P$) in the low temperature range (the orange solid line is a fit with $C_P(T) = \beta T^3 + \delta T^2$ (eqn. S2) explained in supplementary material suggesting A-type or canted antiferromagnetic ordering at these temperatures) and right inset is to clearly show the response at $T_S \sim 330$ K.*

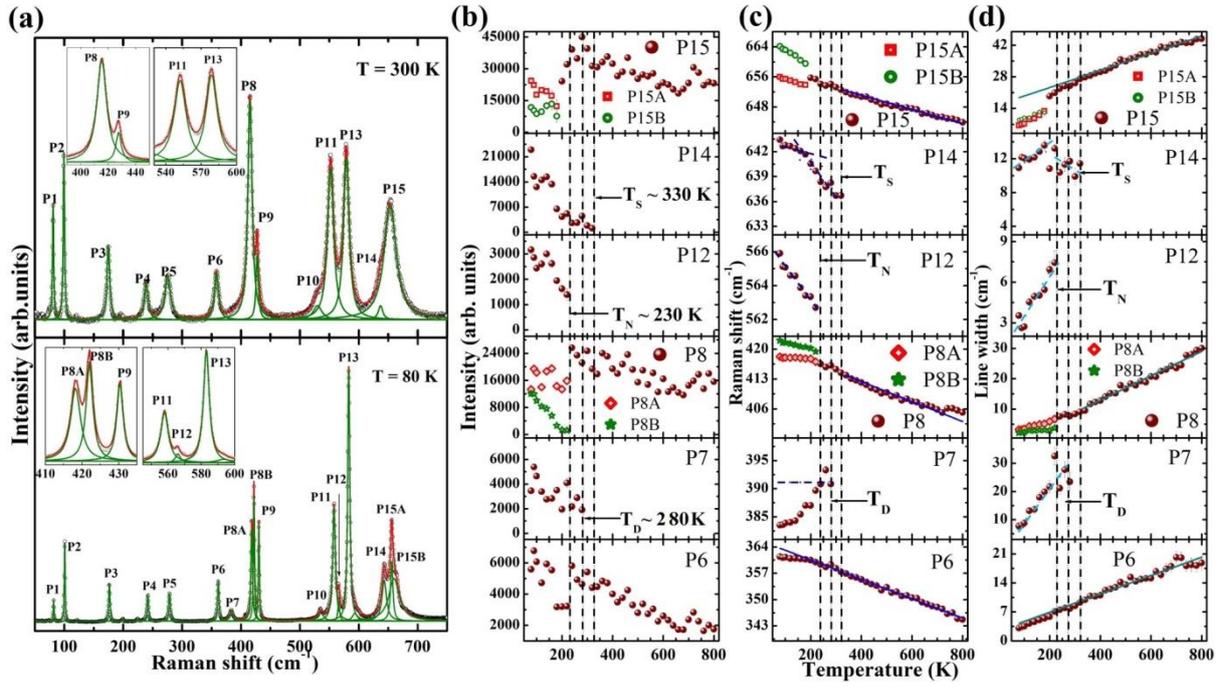

*Figure 3. (Colour online) (a) Raman spectra of 15R-BaMnO$_3$ collected at 80 K and 300 K, (b) spectral weight (intensity), (c) frequency, and (d) linewidth of selected phonons are plotted against temperature. Solid lines in (c) and (d) represent the fitting with anharmonic model (Eq. 2 and 3 in text) whereas the dashed lines for modes P12 and P14 below $T_N$ are guide to eye. Error bars are smaller than or comparable to the symbol size.*

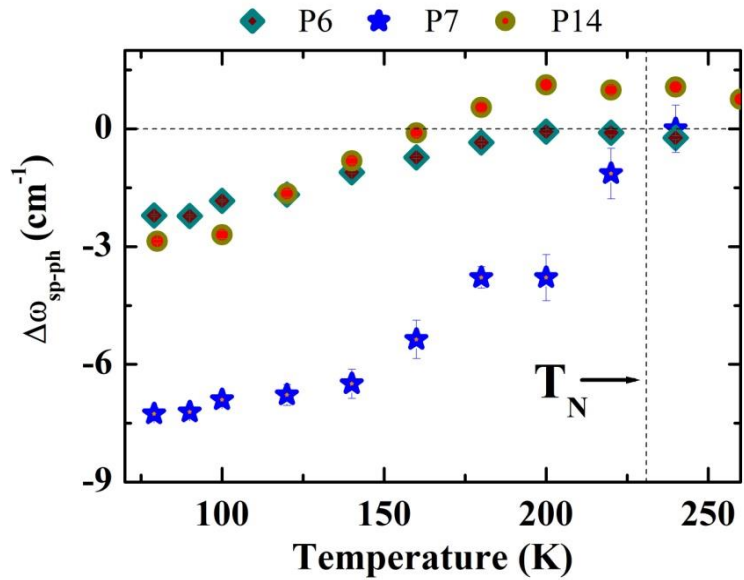

*Figure 4. (Colour online) (a) The deviation of phonon frequency from anharmonic behaviour ($\Delta\omega_{sp-ph}$) as a function of temperature (see text). Error bars are within or comparable to the symbol size.*

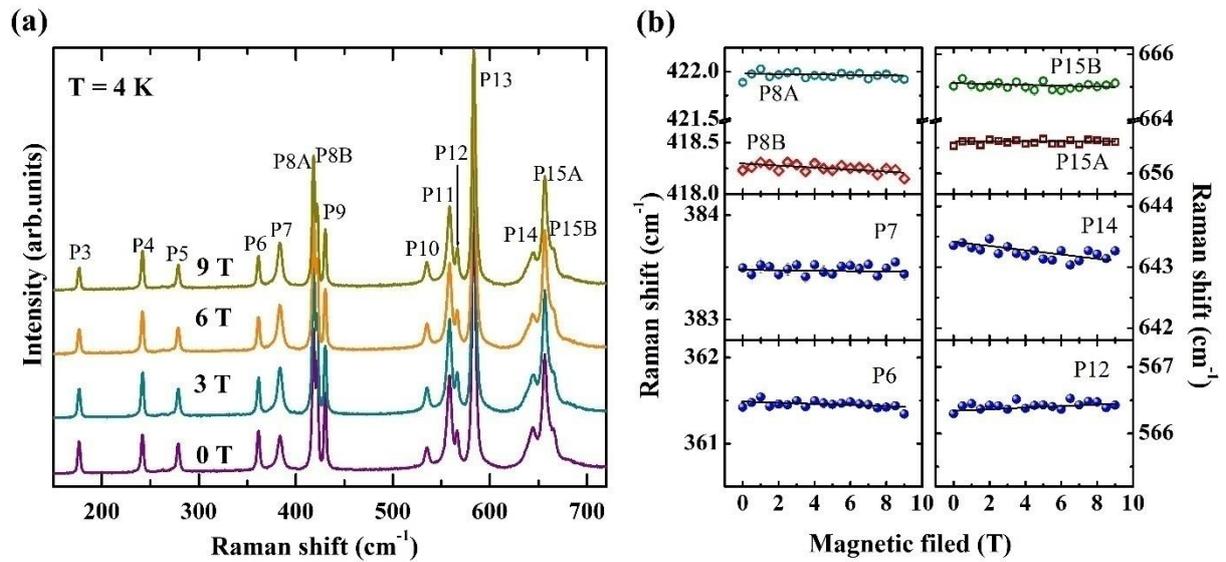

*Figure 5. (Colour online) (a) Raman spectra of 15R-BaMnO$_3$ collected under a few applied magnetic fields at 4 K, (b) frequency of selected phonons are plotted as a function of magnetic field, solid lines are guide to eye showing a decreasing trend with increasing field. Error bars are smaller than the symbol size.*

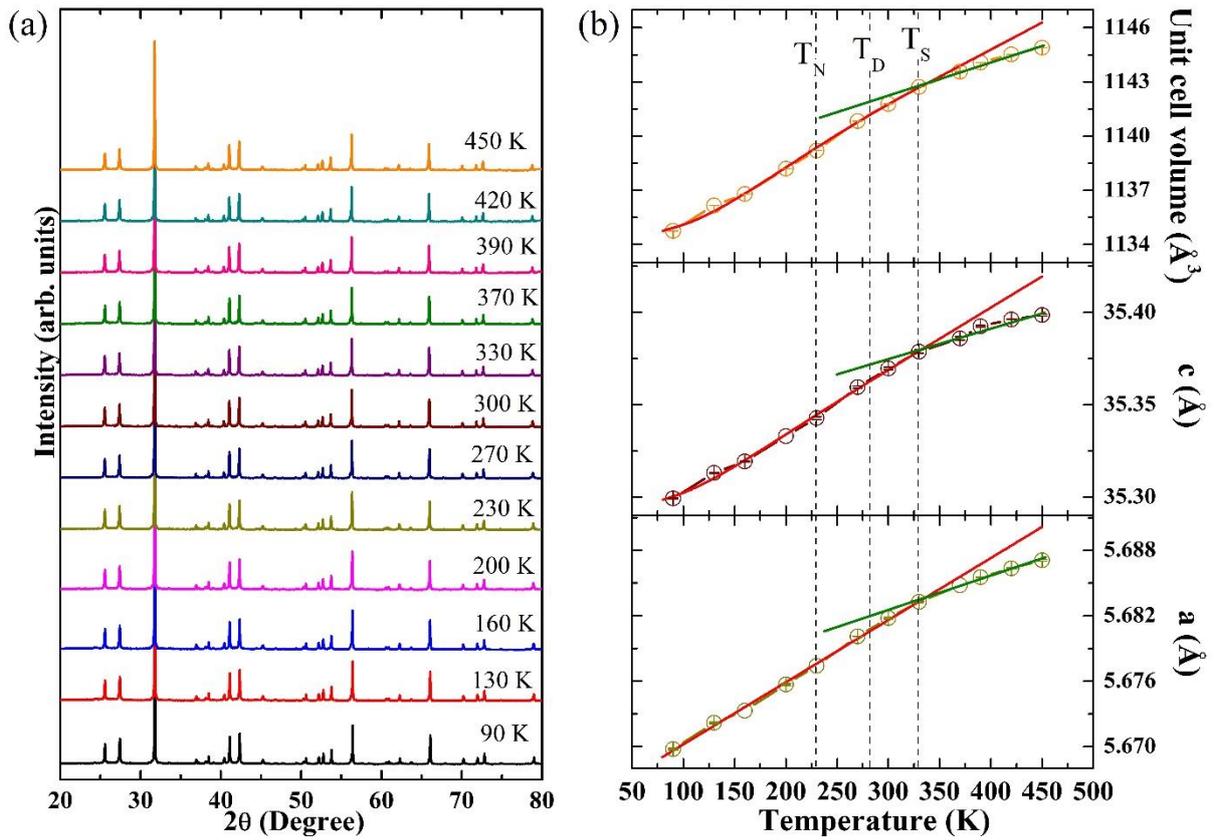

*Figure 6. (Colour online) (a) X-ray diffraction patterns collected at a few temperatures and (b) unit cell parameters as a function of temperature, red solid lines represent fittings presenting the expansion (eq. 6 and 7 explained in text), green solid lines are also fit with Eq. 6 and 7 showing the deviation from expected behaviour. Error bars in (b) are within the symbol size.*

# Supplemental material

# Signatures of magnetostriction and spin-phonon coupling in magnetoelectric hexagonal 15R-BaMnO$_3$

Bommareddy Poojitha, Anjali Rathore, Ankit Kumar, and Surajit Saha*

*Department of Physics, Indian Institute of Science Education and Research, Bhopal 462066, India*

*\*Correspondence: surajit@iiserb.ac.in*

This supplemental material file contains additional data and information about the magnetic and heat capacity measurements, assignment of modes, polarization-, temperature-, and magnetic field-dependent Raman spectra, and powder XRD data. The figures and respective details are given below.

**SM1: Structural details from x-ray diffraction**

The Wyckoff positions for each atom extracted from Rietveld refinement are listed in Table S1. Results indicate the oxygen vacancies at O2 site. The amount of oxygen deficiency (~ 0.01) is similar to the value expected for 15R structure [1].

Table S1. Rietveld refined Wyckoff positions of atoms in 15R-BaMnO$_3$ unit cell.

| atom | site | x | y | z | F.O |
|------|------|---|---|---|-----|
| Ba1 | 3a | 0 | 0 | ½ | 1 |
| Ba2 | 6c | 0 | 0 | 0.36611 | 1 |
| Ba3 | 6c | 0 | 0 | 0.23464 | 1 |
| Mn1 | 3a | 0 | 0 | 0 | 1 |
| Mn2 | 6c | 0 | 0 | 0.13843 | 1 |
| Mn3 | 6c | 0 | 0 | 0.06836 | 1 |
| O1 | 9d | ½ | 0 | ½ | 1 |
| O2 | 18h | 0.51947 | 0.48053 | 0.23094 | 0.9875(1) |
| O3 | 18h | 0.51573 | 0.48427 | 0.36708 | 1 |

## SM2: Magnetism and Specific heat

Inverse susceptibility as a function of temperature of 15R-BaMnO$_3$ is shown in Figure S1 that exhibits non-Curie-Weiss behaviour above T$_S$; where, M(T) increases (inverse susceptibility decreases) slightly with increasing temperature till 400 K, implying that there may be short-range spin-spin correlations even in the paramagnetic phase. As the temperature is lowered below T$_S$, ZFC and FC curves in M(T) meet down to T$_N$ and show significant bifurcation below T$_N$, especially below T$^*$ ~ 43 (± 0.2) K exhibiting a peak-like feature in ZFC curve. These signatures suggest that 15R-BaMnO$_3$ possibly undergoes a phase transition from antiferromagnetic (AFM) state to ferrimagnetic or canted AFM or a glassy state. The spin structure in antiferromagnetic phase is shown in Figure S2. To get further clarification about the nature of the observed magnetic transitions, we have collected magnetic hysteresis M(H) loops at a few typical temperatures and shown in Figure S3. The isothermal magnetization *versus* magnetic-field curves are linear passing through the origin throughout the temperature range down to 43 K which confirms the antiferromagnetic ordering in the system below T$_S$. Notably, clear hysteresis loops are observed below T$^*$. The coercive field (H$_C$) is found to decrease with increasing temperature and it vanishes above T$^*$~ 43 K (inset of Figure S3). To be noted that the M(H) loops do not saturate even up to 7 Tesla of applied magnetic field which suggests that the ordering below T$^*$ is either ferrimagnetic or canted AFM or spin glass. Importantly, a broad peak feature is observed in the first derivative of specific heat with respect to temperature thus suggesting that the magnetic ordering below T* is of canted antiferromagnetic type (Figure S4) [2].

In general, the temperature-dependent heat capacity can be written as;

$$C(T) = C_{elect} + C_{latt} + C_{mag} \quad (S1)$$

The term $C_{elect} = \gamma T$ is the electronic contribution, $C_{latt} = \beta T^3 + \alpha T^5$ is the lattice contribution, while the $C_{mag}$ is the magnetic contribution to the specific heat. $C_{mag} = \delta T^{3/2}$ in case of a ferromagnet. For a standard Néel antiferromagnetically ordered phase, the magnetic contribution is proportional to $T^3$ which is similar to the Debye lattice term. However, for A-type antiferromagnetic order it can be written as $C_{mag} = \delta T^2$ [3,4]. The specific heat at low temperatures (below 22 K) is non-linear in temperature thus ruling out any electronic contribution due to the insulating nature of 15R-BaMnO$_3$.

As can be seen in Figure 2b (left inset), the heat capacity data of 15R-BaMnO$_3$ in the low temperature region (below 22 K) is analysed with the equation;
$$C(T) = \beta T^3 + \delta T^2 \quad (S2)$$

A good fitting is obtained with R$^2$ = 0.996 upon considering Eq. (S2). The $\delta T^2$ term is considered due to canted antiferromagnetic nature of BaMnO$_3$ in the low temperature region. An addition of $T^{3/2}$ term does not appropriately fit the experimental data while addition of $\alpha T^5$ term does not improve the fitting either thus ruling out the contributions relevant to these terms. To be noted that the $T^3$ term is dominant below about 20 K, as shown in Figure S4b. The obtained fitting parameters are $\beta = 0.21 \pm 0.02 \frac{mJ}{mol.K^4}$ and $\delta = 2.45 \pm 0.60 \frac{mJ}{mol.K^3}$. The Debye temperature is extracted from $\beta$ value using the relation $\theta_D = \frac{12\pi^4 pR}{5\beta}$; where $p$ is number of atoms per formula unit, and R is universal gas constant. The obtained value for Debye temperature is $\theta_D \sim 358 \pm 34\ K$ which is in very good agreement with the value obtained from the temperature dependent x-ray diffraction results ($\theta \sim 352 \pm 20\ K$).

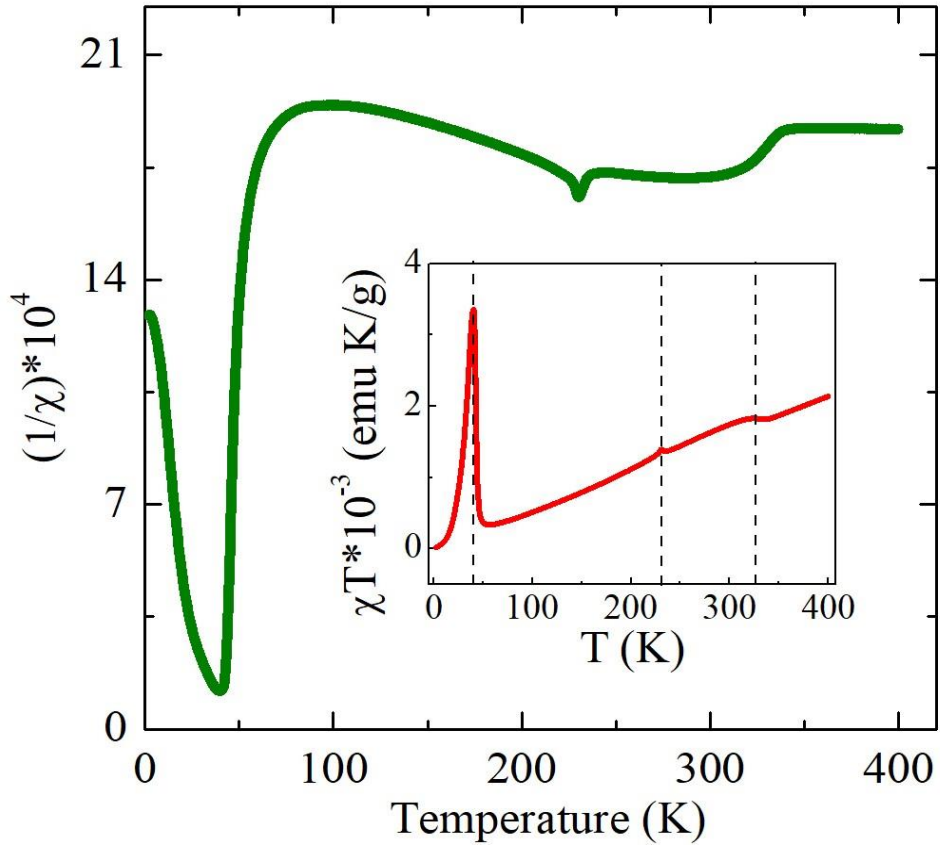

Figure S1. $\chi^{-1}$ vs T, inset shows the plot of $\chi T$ vs T for 15R-BaMnO$_3$.

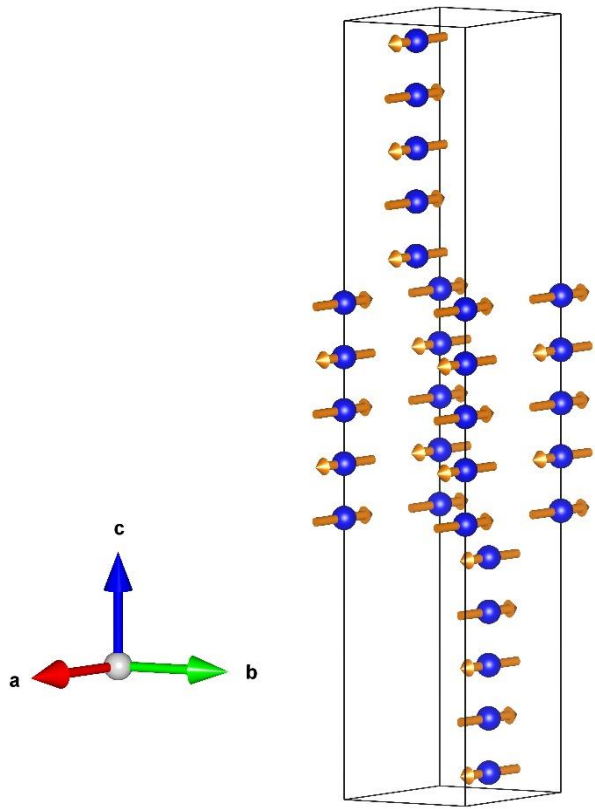

Figure S2. Spin structure in antiferromagnetic phase (below $T_N \sim 230$ K) for 15R-BaMnO$_3$.

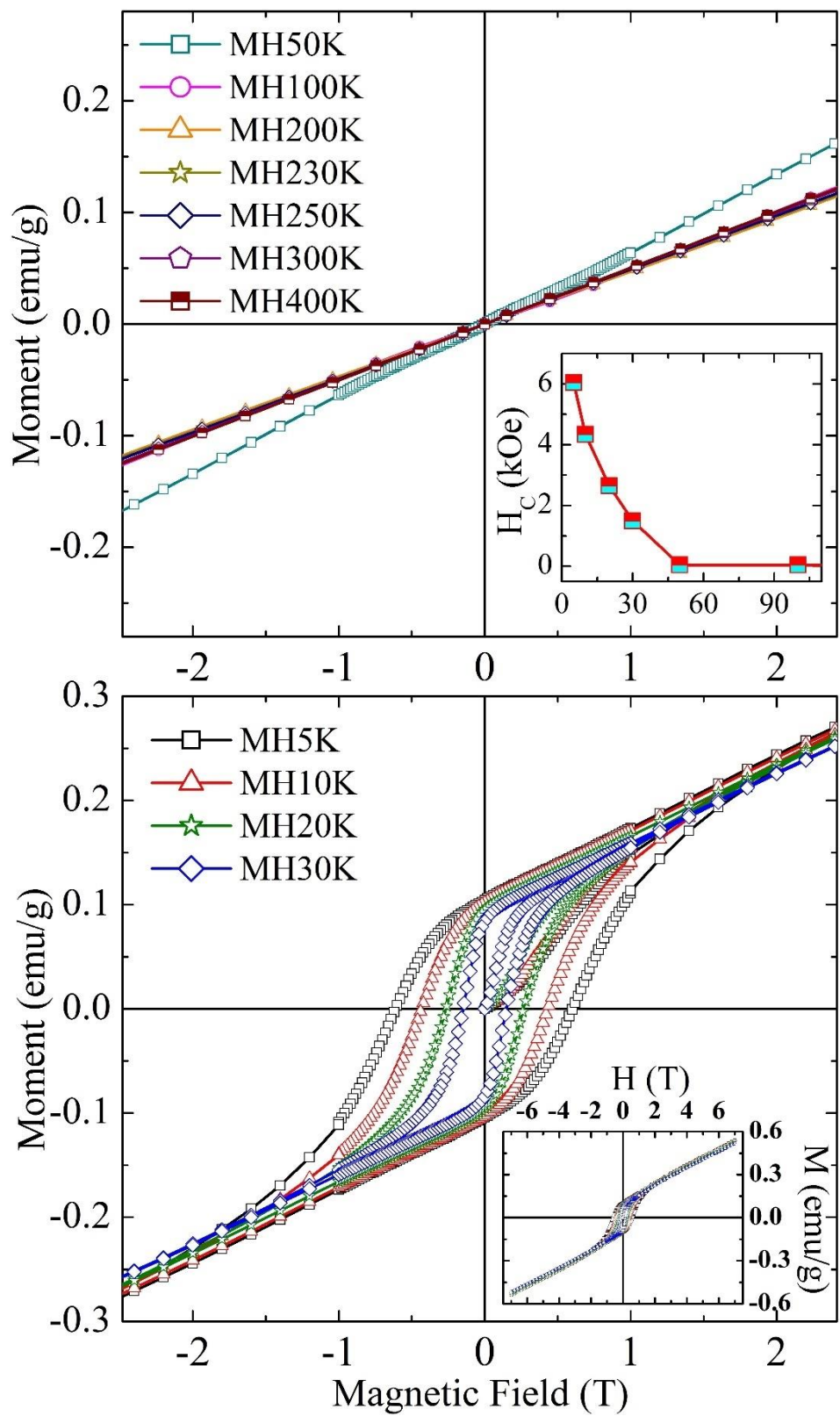

Figure S3. MH curves at a few typical temperatures. Inset in the top panel shows the plot of coercivity as a function of temperature. While the inset in the bottom panel gives the M *vs* H up to the highest field (± 7 T) measured.

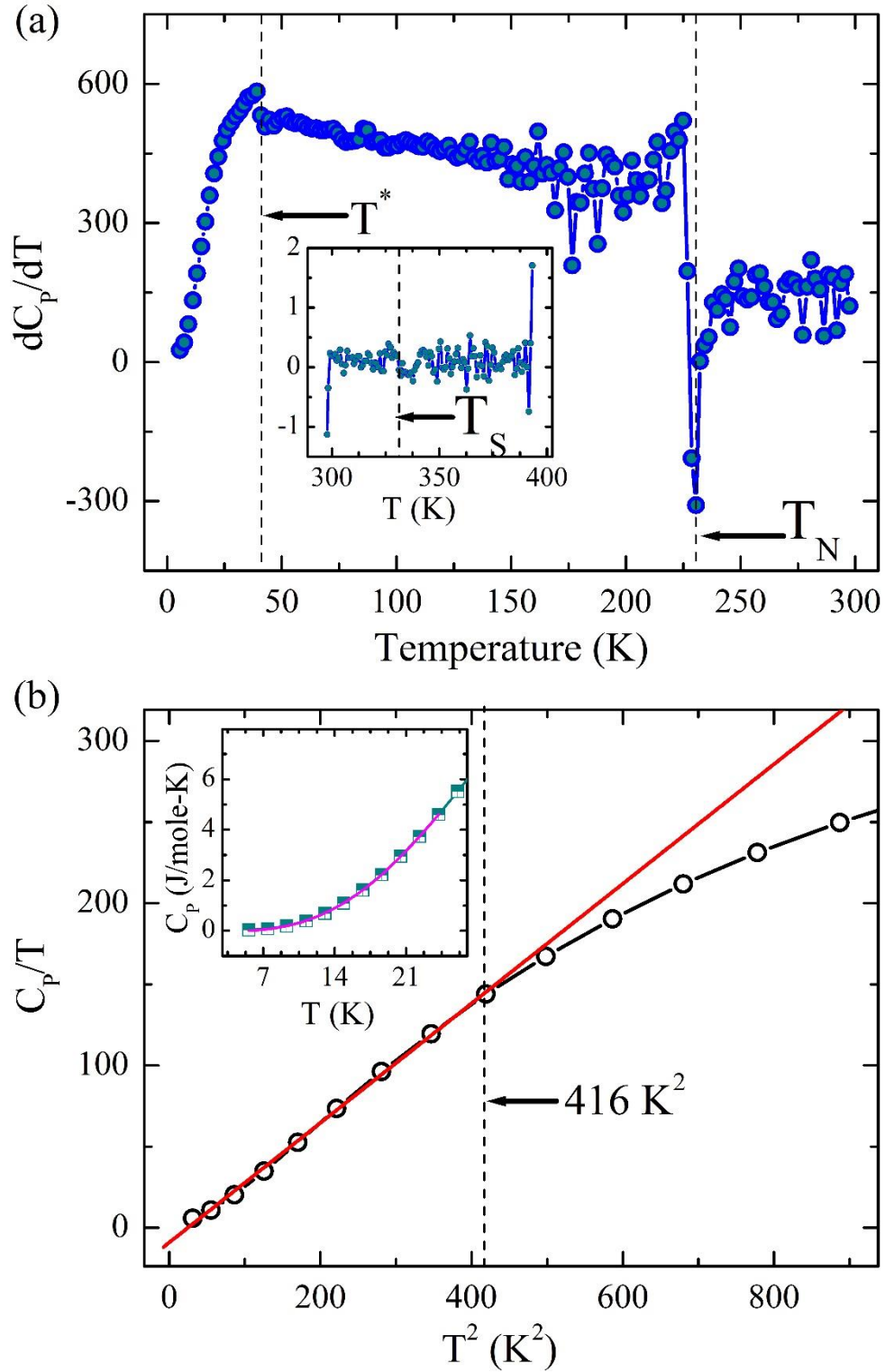

Figure S4. (a) First derivative of specific heat ($dC_P/dT$) is plotted as a function of temperature, (b) $C_P/T$ vs T plot to show the $T^3$ dependence of specific heat in the low temperature region (T<20 K). In the inset, the solid line is a fit with Eq. S2 explained in the text.

**SM3: Symmetry assignment for Raman modes**

Raman spectra are collected in two polarization configurations as shown in Figure S5. The laser from the source is linearly polarized along Y, the collected spectrum in the same direction of polarization is indicated as YY (parallel) polarization and collected in orthogonal direction of polarization is YX (perpendicular) polarization. The symmetry for Raman active modes is assigned based on group theoretical prediction and polarization-dependent Raman spectra. Ideally it is expected that the modes with $A_g$ symmetry would completely get suppressed under cross polarization and $E_g$ modes appear in both parallel and cross polarization configurations. But we have seen only slight suppression/changes in the intensity rather than mode disappearing which is due to the polycrystalline nature of the sample (where crystalline directions are not well defined) and possibly due to polarization leakage, arising from the breaking of Raman selection rules [5].

Note that the P7 mode is present in polarization-dependent Raman collected at room temperature whereas it disappears above 280 K in temperature-dependent measurements. It is due to the fact that the polarization measurements are performed outside the cryostat where the room temperature is close to 285 K. According to the harmonic approximation, the phonon frequency can be expressed as $\omega \propto \sqrt{\frac{k}{m}}$. The phonon frequency is less for the vibrations involving heavier atoms and relatively high for modes associated with lighter atoms. The group theory predicts 4, 4, and 10 Raman active vibrations involving Ba, Mn, and O atoms, respectively, for the 15R phase (R-3m symmetry) as given in Table S2. The modes below 175 cm$^{-1}$ are associated with Ba displacements. The phonons of frequency between 190 and 400 cm$^{-1}$ arises from Mn vibrations. High frequency modes are related to the displacement of O atoms. The frequency of modes at room temperature falls at 80.9($\pm$ 0.1) [$E_g$], 99.3 ($\pm$ 0.1) [$A_{1g}$], 174.7 ($\pm$ 0.1) [$E_g$], 238.9 ($\pm$ 0.1) [$E_g$], 275.6 ($\pm$ 0.1) [$E_g$], 358.4 ($\pm$ 0.1) [$E_g$], 415.3($\pm$ 0.1) [$E_g$], 427.5($\pm$ 0.1) [$E_g$], 530.0($\pm$ 0.4) [$A_{1g}$], 552.2($\pm$ 0.1) [$E_g$], 578.8($\pm$ 0.1) [$E_g$], 636.9($\pm$ 0.3) [$A_{1g}$], 653.3($\pm$ 0.1) [$E_g$] cm$^{-1}$. The weak modes of frequency 120.6 ($\pm$ 0.2) [$A_{1g}$] and 195.9 ($\pm$ 0.2) [$A_{1g}$] cm$^{-1}$ are shown as inset of Figure S5. The symmetry of the phonon is written in square brackets.

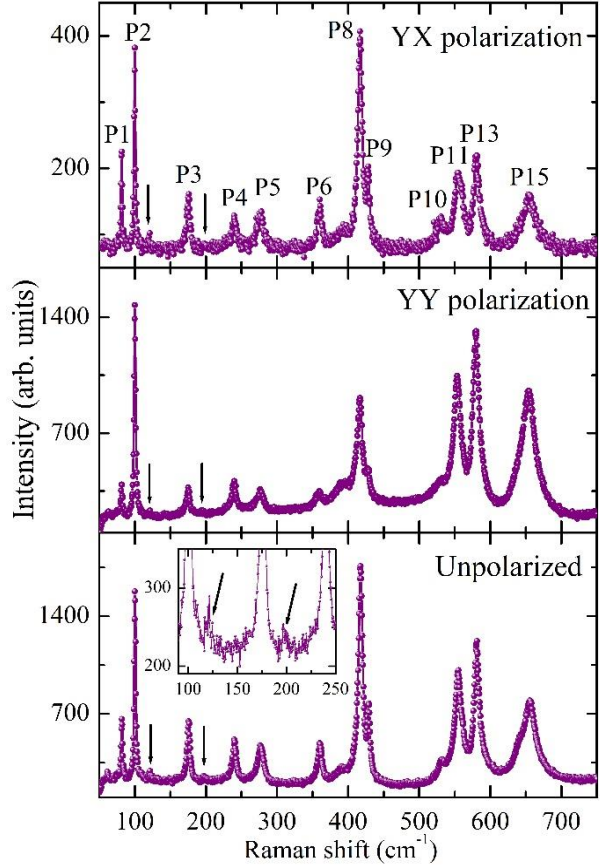

Figure S5. Raman spectra collected in parallel and perpendicular polarization configurations. Inset is enlarged view to show the weak modes at ~ 120 and ~ 196 cm$^{-1}$.

Table S2. Mode activity for phonons in 15R-BaMnO$_3$.

| Space group: R-3m (No. 166) | |
|---|---|
| It has 15 formula units (75 atoms) per unit cell | |
| Five formula units (25 atoms) per primitive cell | |
| Wyckoff positions | Irreducible representations |
| Ba1: 3b | $A_{2u} + E_u$ |
| Ba2: 6c | $A_{2u} + E_u + A_{1g} + E_g$ |
| Ba3: 6c | $A_{2u} + E_u + A_{1g} + E_g$ |
| Mn1: 3a | $A_{2u} + E_u$ |
| Mn2: 6c | $A_{2u} + E_u + A_{1g} + E_g$ |
| Mn3: 6c | $A_{2u} + E_u + A_{1g} + E_g$ |
| O1: 9d | |
| O2: 18h | $2A_{2u} + 3E_u$ |
| O3: 18h | $2A_{2u} + 3E_u + 2A_{1g} + 3E_g$ |
| | $2A_{2u} + 3E_u + 2A_{1g} + 3E_g$ |
| Total irreducible representations $\Gamma_{Raman} = 8A_{1g} + 10E_g$ and $\Gamma_{IR} = 12A_{2u} + 15E_u$ | |

**SM4: Temperature-dependent Raman spectra**

The Raman spectra at a few typical temperatures are stacked in Figure S6. The fitting parameters extracted from cubic anharmonic equations for temperature-dependent frequency and linewidth are listed in Table S3. Phonon parameters (intensity, frequency, and linewidth) for P1-P5, P9, P11, and P13 modes are plotted as a function of temperature in Figure S7-S9. The finite deviation from the expected anharmonic behaviour is observed for all modes which can be attributed to magnetostriction in the system.

Table S3. Anharmonic fitting (Eq. 2 and 3 in main text) parameters for Raman active phonons in 15R-BaMnO$_3$.

| Mode | Temperature range where the mode exists | $\omega_0$ (cm$^{-1}$) | A | $\Gamma_0$ (cm$^{-1}$) | C |
|---|---|---|---|---|---|
| P1 | 80-800 K | 82.42 ± 0.06 | -0.030 ± 0.001 | 1.30 ± 0.06 | 0.020 ± 0.001 |
| P2 | 80-800 K | 100.90 ± 0.06 | -0.030 ± 0.001 | 1.50 ± 0.06 | 0.020 ± 0.001 |
| P3 | 80-800 K | 177.50 ± 0.09 | -0.100 ± 0.002 | 1.80 ± 0.15 | 0.130 ± 0.003 |
| P4 | 80-800 K | 242.10 ± 0.09 | -0.150 ± 0.002 | 2.10 ± 0.24 | 0.22 ± 0.01 |
| P5 | 80-800 K | 279.60 ± 0.10 | -0.240 ± 0.003 | 1.30 ± 0.28 | 0.50 ± 0.01 |
| P6 | 240-800 K | 365.70 ± 0.24 | -0.540 ± 0.009 | 1.60 ± 0.29 | 0.50 ± 0.01 |
| P7 | 80-280 K | Anharmonic equation is not fitted as this mode is no longer purely anharmonic | | | |
| P8 | 240-800 K | 421.80 ± 0.33 | -0.560 ± 0.016 | 0.01 ± 0.00 | 1.00 ± 0.02 |
| P9 | 80-800 K | 431.38 ± 0.11 | -0.330 ± 0.006 | 0.84 ± 0.15 | 0.32 ± 0.01 |

| P10 | 80-800 K | 536.54 ± 0.55 | -0.510 ± 0.035 | 8.38 ± 2.09 | 1.67 ± 0.13 |
| P11 | 80-800 K | 561.10 ± 0.22 | -1.030 ± 0.015 | 3.33 ± 0.38 | 0.89 ± 0.02 |
| P12 | 80-220 K | Anharmonic equation is not fitted as this mode is no longer purely anharmonic | | | |
| P13 | 80-800 K | 584.71 ± 0.13 | -0.690 ± 0.009 | 2.12 ± 0.41 | 0.95 ± 0.03 |
| P14 | 80-330 K | Anharmonic equation is not fitted as this mode is no longer purely anharmonic | | | |
| P15 | 240-800 K | 658.68 ± 0.24 | -0.710 ± 0.017 | 13.10 ± 0.55 | 1.56 ± 0.04 |

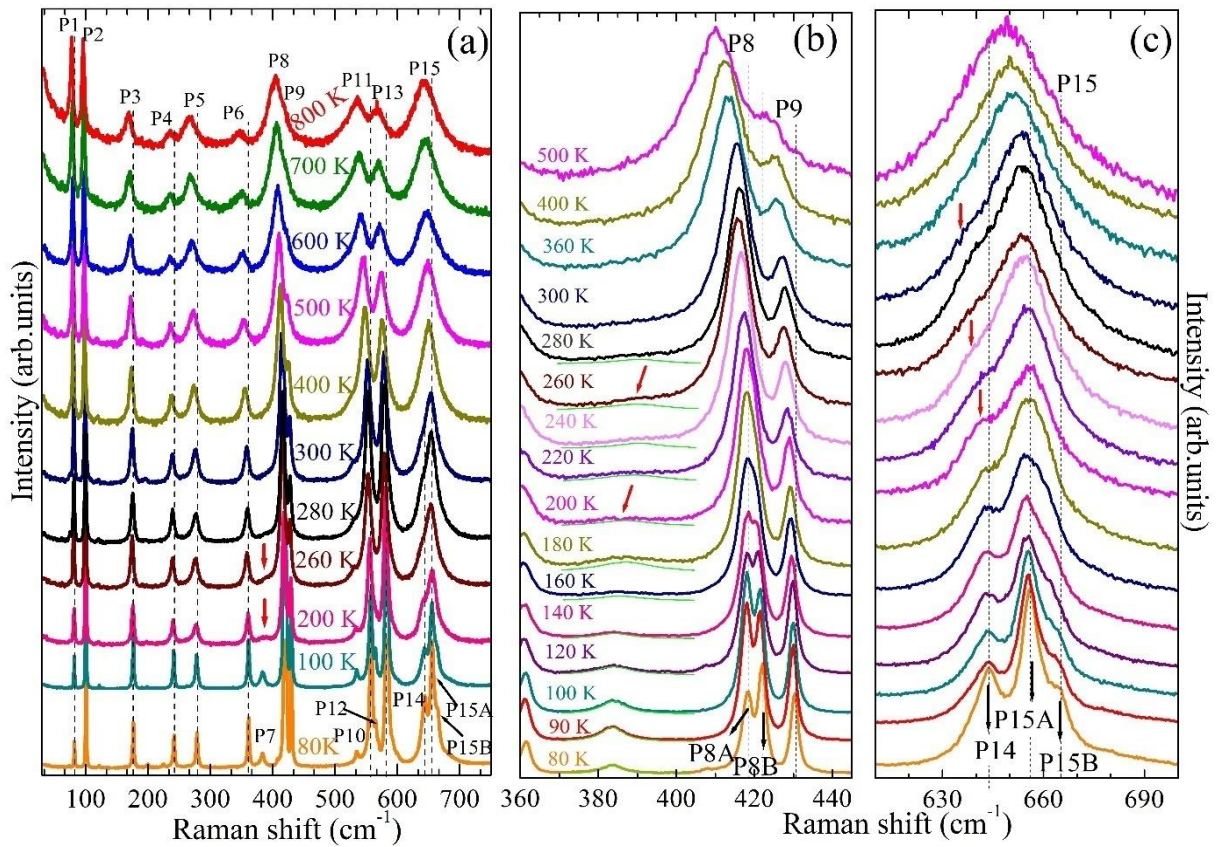

Figure S6. (a) Raman spectra of 15R-BaMnO$_3$ collected at a few temperatures where P7 disappears above ~ 280 K (T$_D$), (b) and (c) show the enlarged view to show the splitting of P8 and P15 phonons, respectively below magnetic ordering temperature (T$_N$).

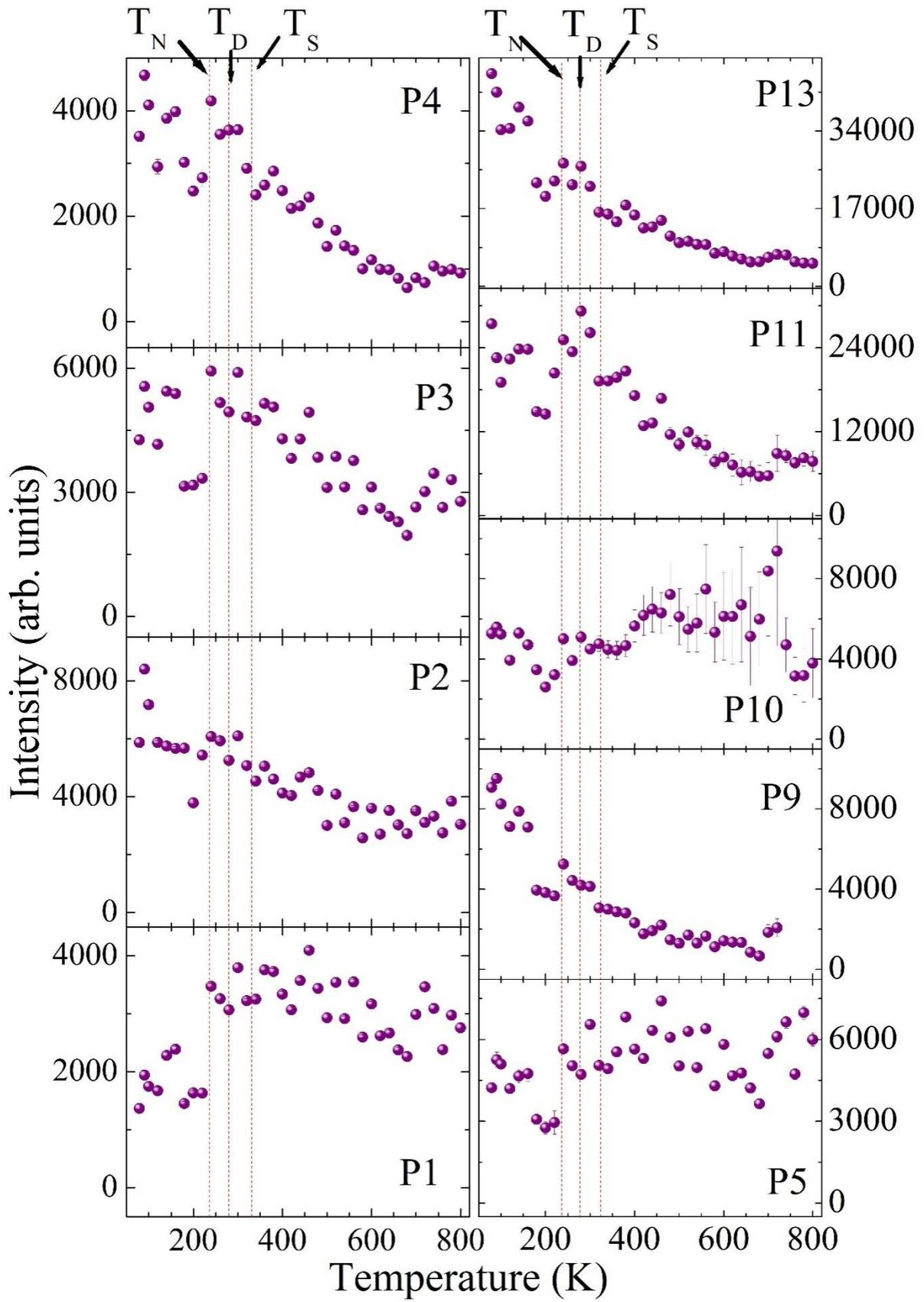

Figure S7. Spectral weight of a few Raman active phonons as a function of temperature.

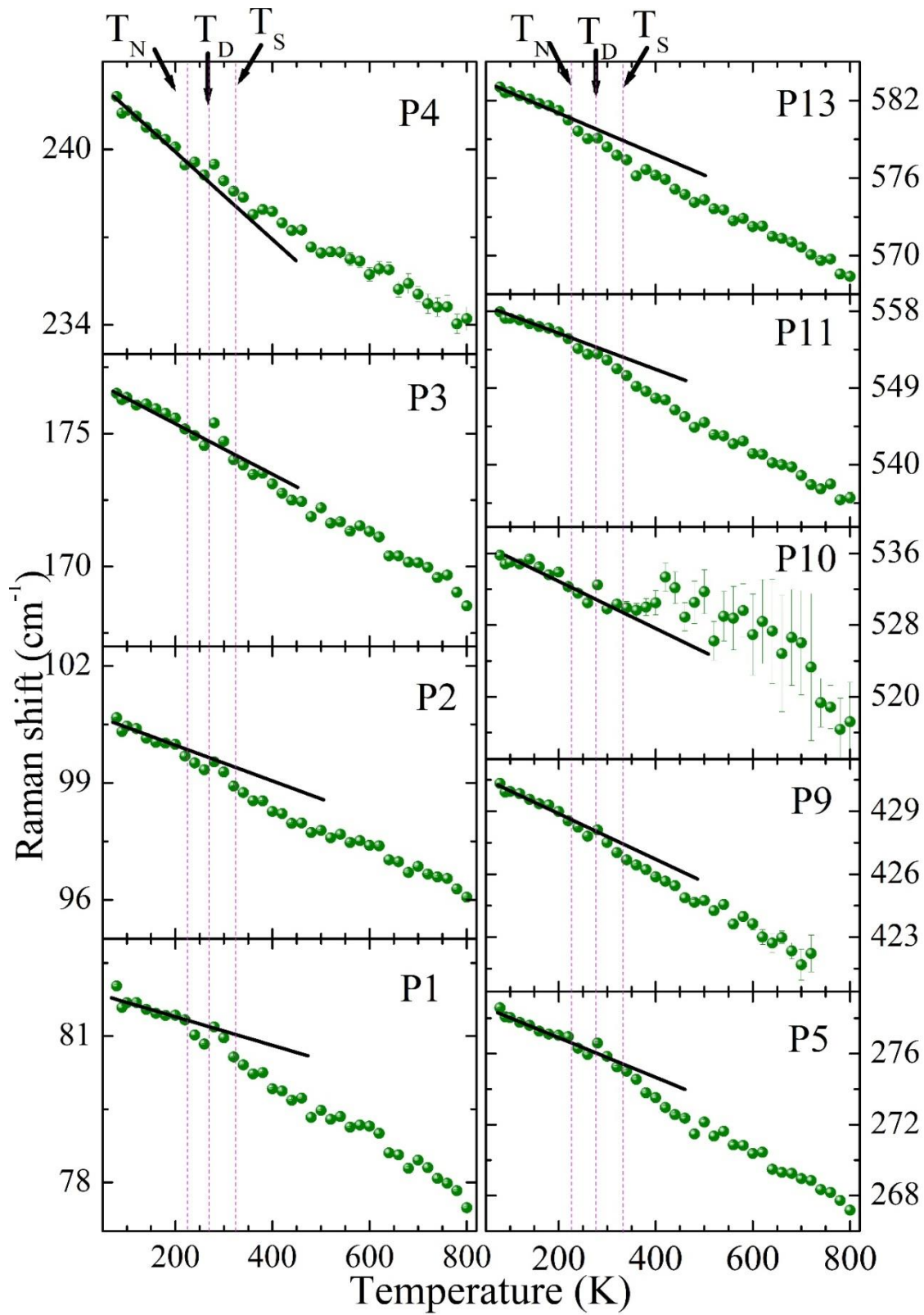

Figure S8. Frequency of a few Raman active phonons as a function of temperature. Solid lines are guide to eye showing the deviation at magnetic transition temperatures.

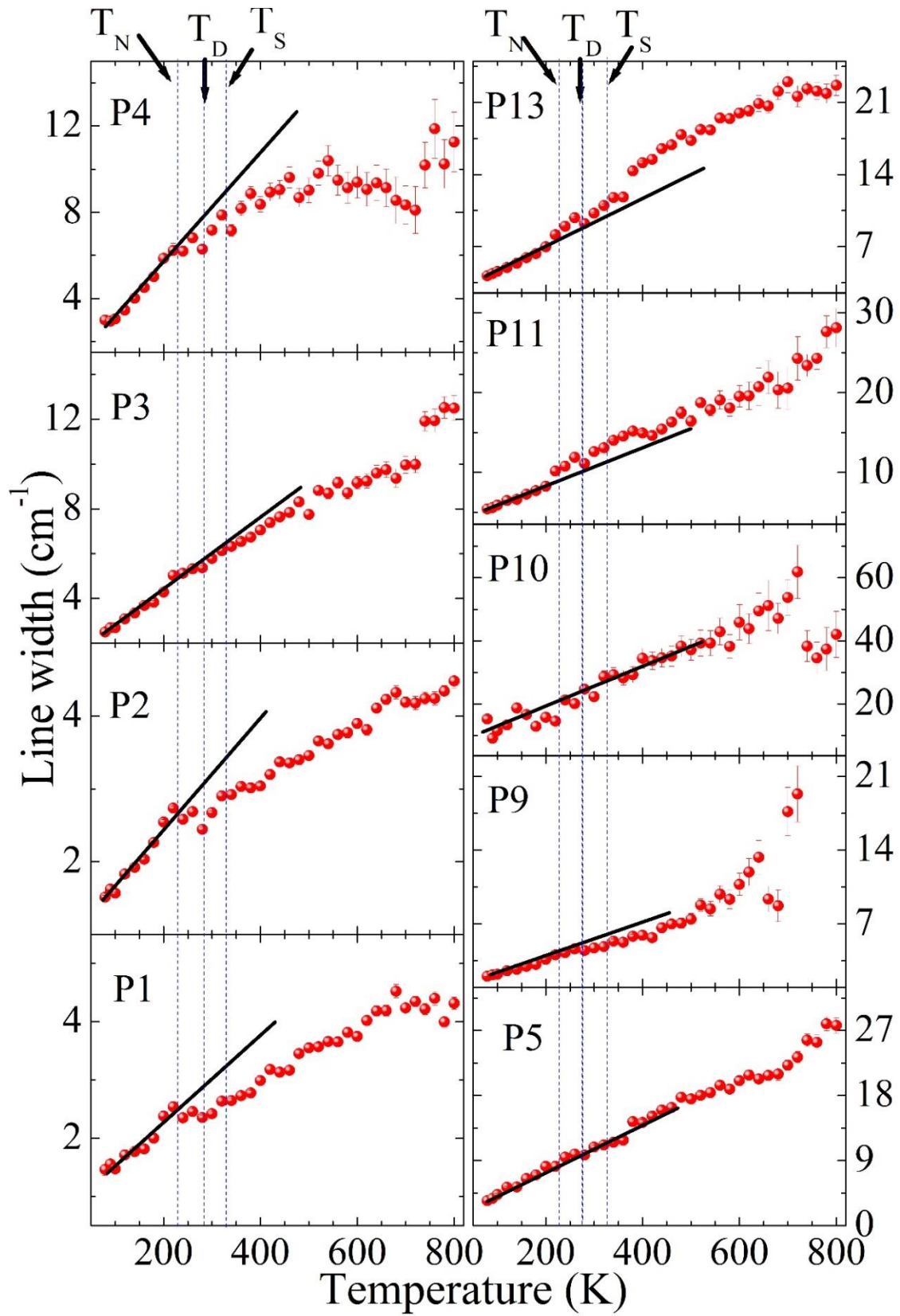

Figure S9. Linewidth of a few Raman active phonons as a function of temperature. Solid lines are guide to eye showing the deviation at magnetic transition temperatures.

## SM5: Spin-phonon coupling

We have analysed the temperature-dependent phonon frequency using Eq. 2 (explained in main text). The deviation from the anharmonic equation (Eq. 2) is calculated by the equation:
$$\Delta\omega(T) = \omega(T) - \omega_{anh}(T)$$
In magnetic materials, the phonon parameters such as frequency and linewidth may respond to the magnetic transition temperature ($T_N$) due to spin-phonon coupling. The change in phonon frequency due to spin-phonon coupling ($\lambda$) in magnetically ordered phase can be written as:
$$\omega(T) = \omega_{anh}(T) - \lambda <S_i.S_j>$$
where $<S_i.S_j>$ defines the spin-spin correlation. The deviation in the phonon frequency from the expected anharmonic behaviour in the magnetically ordered phase is:
$$\Delta\omega(T) = \Delta\omega_{sp-ph}(T) = \omega(T) - \omega_{anh}(T) = -\lambda <S_i.S_j>$$
We have estimated the spin-phonon coupling constant using the following relation as explained in the main text:
$$\lambda_{sp} = -\frac{\omega(T_{Low}) - \omega_{anh}(T_{Low})}{[\Phi(T_{Low}) - \Phi(2T_N)]S^2}$$
where $\omega(T_{Low})$ is the experimental phonon frequency at the lowest temperature (T ~ 80 K) recorded while the $\omega_{anh}(T_{Low})$ is the anharmonic estimate of the phonon frequency at the same temperature. The spin (S) is 3/2 for $Mn^{+4}$. $\Phi$ is the short-range order parameter. Since P7 phonon does not exist in paramagnetic phase, in the absence of experimental data with anharmonic behaviour, we have used $\omega$ at $T_N$ ($\omega_{T_N}$) in the above equation for $\omega_{anh}(T_{Low})$, to estimate $\lambda_{sp}$. This procedure will result in underestimation of the $\lambda_{sp}$ for P7.

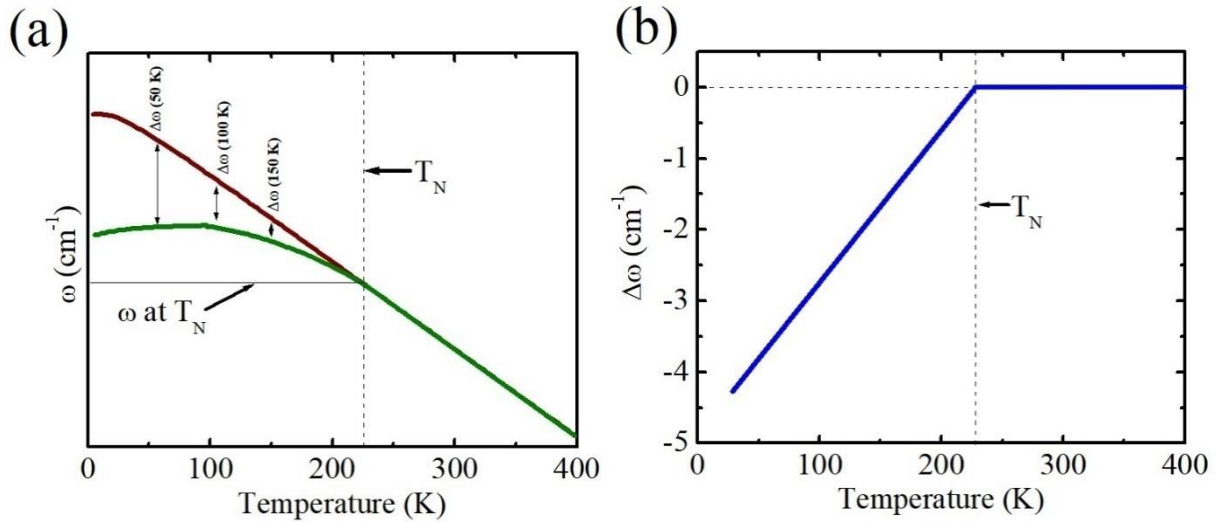

Figure S10. The schematics for (a) $\omega$ vs $T$ showing the deviation from the expected anharmonic trend, and (b) $\Delta\omega_{sp-ph}(T)$ vs $T$.

**SM6: Magnetic field dependent Raman spectra**

Figure S11a shows the Raman spectra collected at a few magnetic fields. The phonon intensity and linewidth for spin-related modes as a function of temperature are plotted in Figure S11b. Intensity, frequency, and linewidth for other modes are plotted in Figures S12, S13, and S14, respectively.

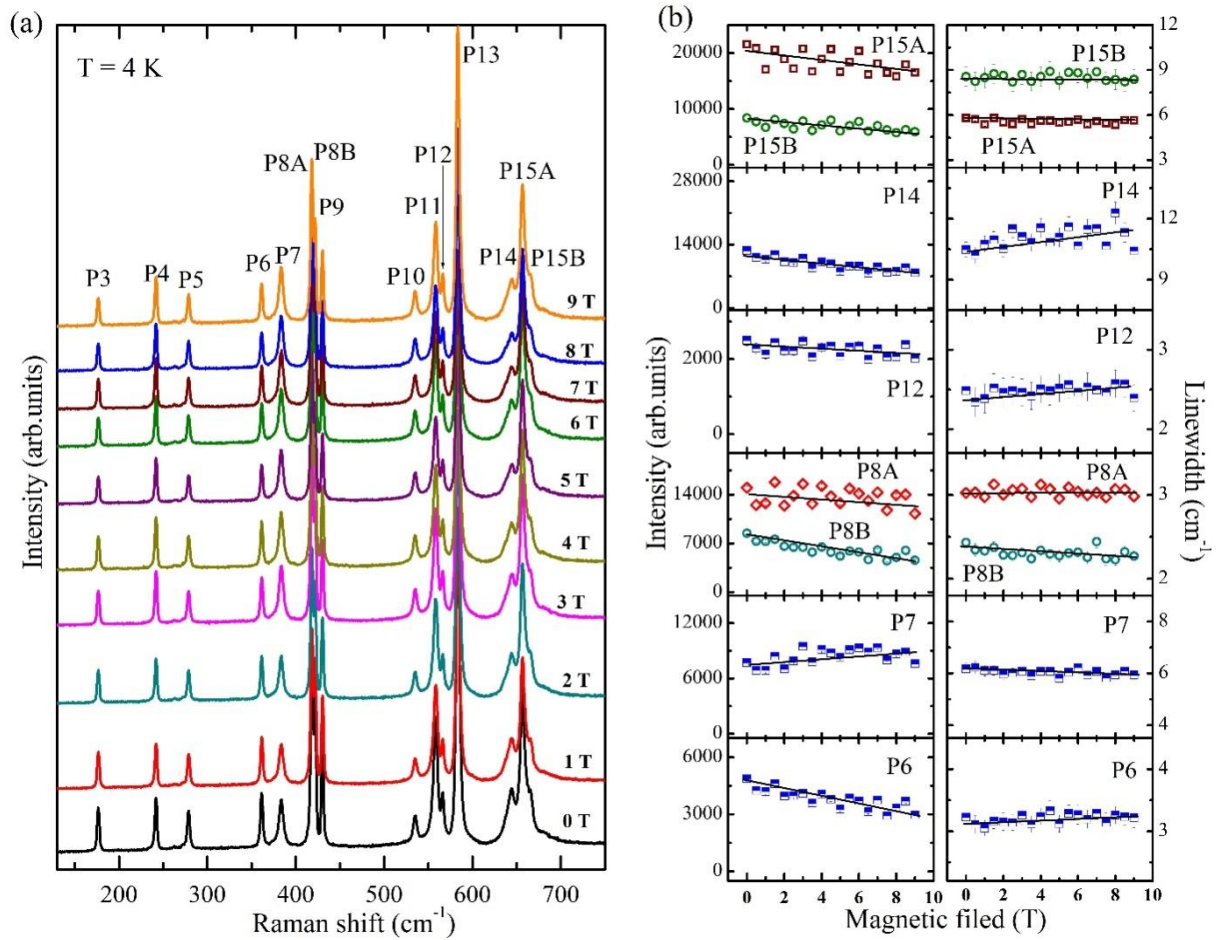

Figure S11. (a) Raman spectra of 15R-BaMnO$_3$ collected at a few magnetic fields, (b) intensity and linewidth of spin-related modes as a function of magnetic field.

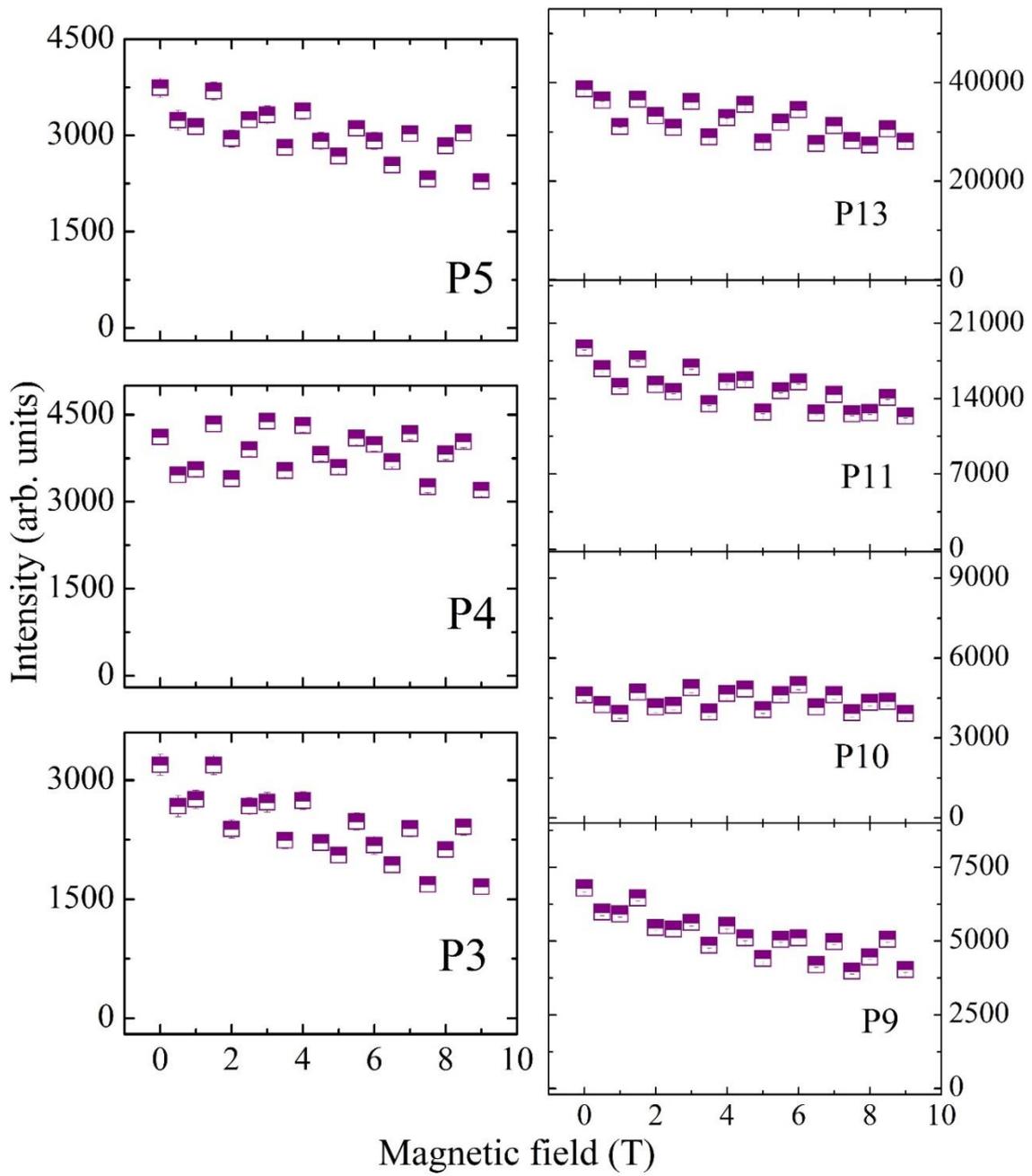

Figure S12. Spectral weight of phonons as a function of applied magnetic field

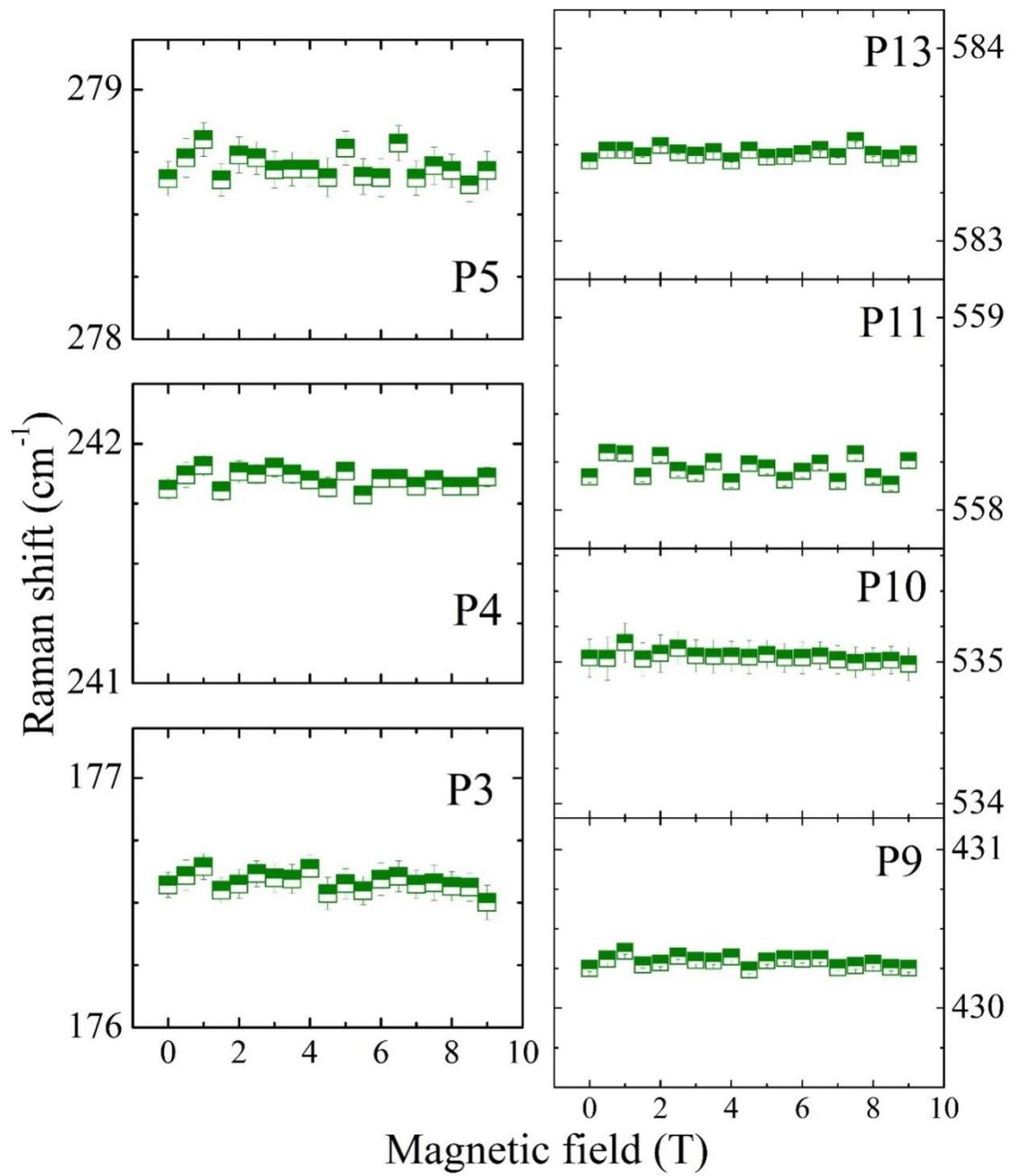

Figure S13. Frequency of phonon as a function of applied magnetic field

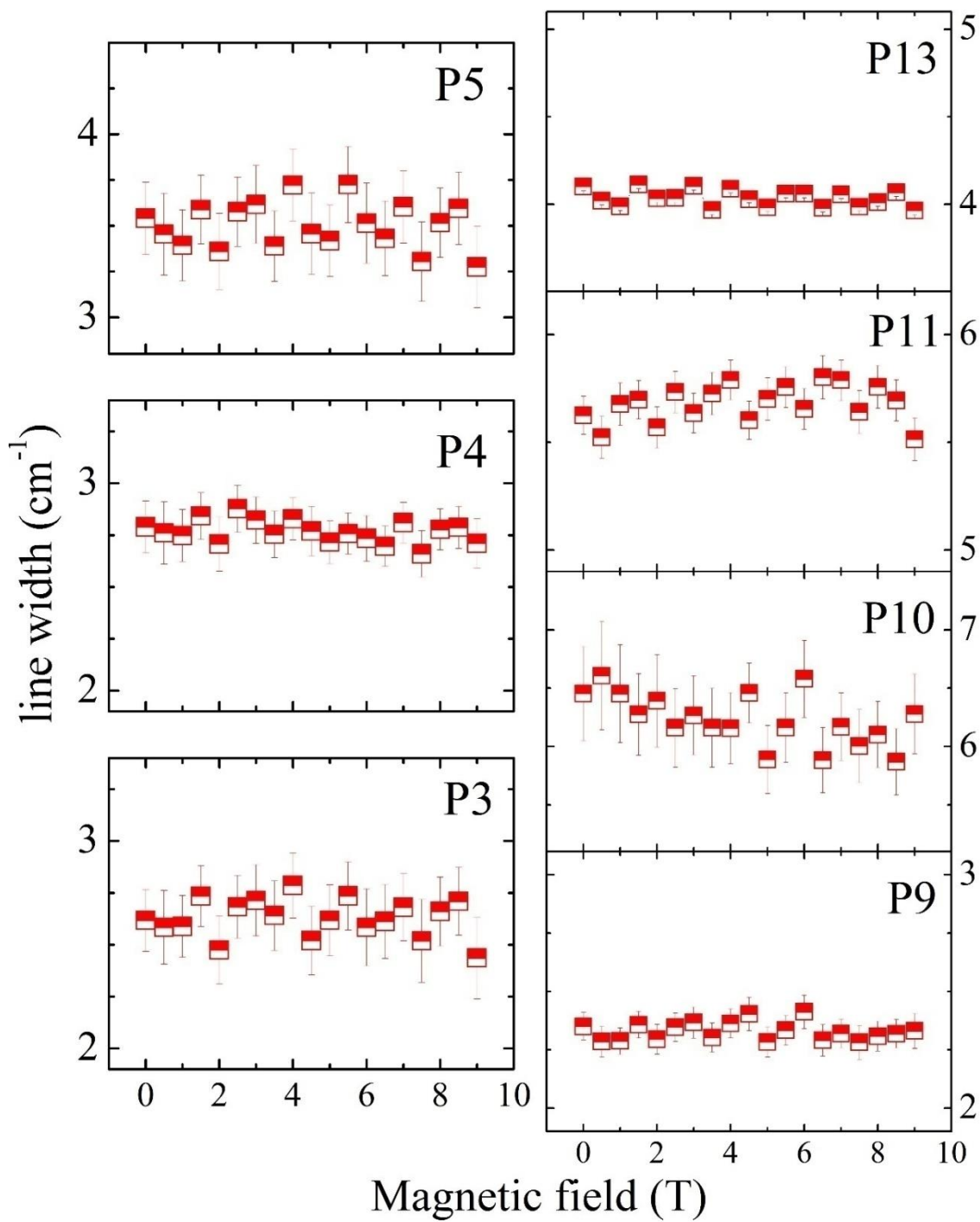

Figure S14. Phonon linewidth as a function of applied magnetic field

**SM7: Thermal expansion: Temperature-dependent x-ray diffraction**

The x-ray diffraction patterns collected at a few typical temperatures (Figure 6a in main text) confirms that the crystal symmetry remains the same throughout the investigated temperature range (90-450 K). The temperature-dependent lattice parameters due to thermal expansion can be written as [6]:

$$a(T) = a_0\left[1 + \frac{be^{\frac{d}{T}}}{T(e^{\frac{d}{T}}-1)^2}\right] \quad \text{and} \quad c(T) = c_0\left[1 + \frac{fe^{\frac{g}{T}}}{T(e^{\frac{g}{T}}-1)^2}\right] \quad (S3)$$

where, $a_0$ and $c_0$ are the in-plane and out-of-plane lattice constants at 0 K, whereas b, d, f, and g are fitting parameters. Unit cell volume expansion with temperature can be expressed as:

$$V(T) = V_0\left[1 + \frac{A}{(e^{\frac{\theta}{T}}-1)}\right] \quad (S4)$$

where, $V_0$ is the cell volume extrapolated to 0 K and $\theta$ is Debye temperature and A is fitting constant. The fitting parameters are given in table S4. The lattice parameters as a function of temperature are shown in Figure 6b in main text. The fitted data in the entire temperature range (90-450 K) does not give reliable values for parameters. The best fitted parameters of temperature-dependent lattice constants with Eq. S3 and S4 are given in Table S3. The fluctuations in lattice parameters around $T_S$ and $T_D$ represent the local modifications in the lattice due to magnetic ordering and the change in electric property. However, unit cell volume fits reasonably well with Eq. (5) in the entire temperature range (90-450 K). The Debye temperature from x-ray diffraction results is $\theta \sim 352 \pm 20$ K. Figure S15 shows the Mn-Mn separation as a function of temperature. All the Mn-Mn bond lengths show an increasing trend with increasing temperature and exhibit a change in slope at $T_S$, similar to the lattice parameters.

Table S4. The best fit parameters of temperature-dependent lattice parameters with Eq. 4 and 5 explained in the main text.

| Lattice parameter (Å) | Selected temperature range (K) | Fitting parameters |
|---|---|---|
| a | 90-450 K | $a_0 = 5.6686 \pm 0.0028$ Å, b = 0.8607 ± 1.2737, and d = 327.12 ± 255.587 |
| | 90-330 K | $a_0 = 5.6692 \pm 0.0005$ Å, b = 1.8253 ± 0.5526, and d = 454.14 ± 64.602 |
| c | 90-450 K | $c_0 = 35.293 \pm 0.0171$ Å, f = 0.8502 ± 1.3009, and g = 336.16 ± 271.17 |
| | 90-330 K | $c_0 = 35.297 \pm 0.0028$ Å, f = 2.0171 ± 0.5038, and g = 488.93 ± 57.128 |
| V | 90-450 K | $V_0 = 1134 \pm 0.25$ Å³, A = 23.76 ± 0.86 and $\theta \sim 352 \pm 20.33\ K$. |
| | 90-330 K | $V_0 = 1134 \pm 0.23$ Å³, A = 27.33 ± 2.22 and $\theta \sim 402 \pm 30.41\ K$. |

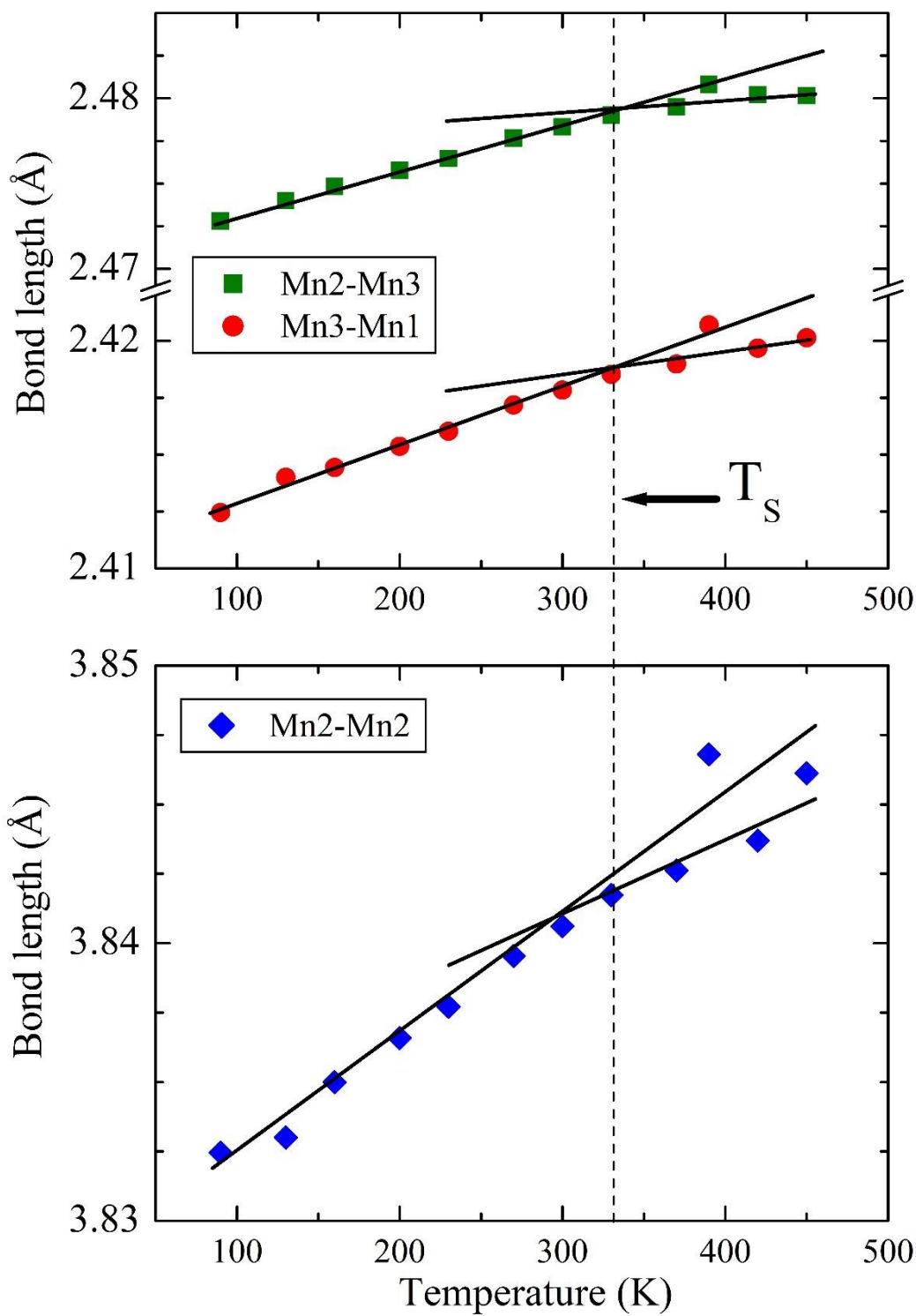

Figure S15: Mn-Mn separation as a function of temperature. Solid lines are guide to eye showing the deviation at $T_S$.

## SM8: Magnetostriction

*Magnetostriction from x-ray diffraction measurements*

The spontaneous volume magnetostriction ($\lambda_{ms}^V$) at a given temperature can be defined as the difference between the volume of the unit cell ($V_{AFM}(T)$) in the magnetically ordered phase (antiferromagnet in 15R-BaMnO₃) at that temperature and the hypothetical volume of the unit cell at the same temperature if it were in the paramagnetic (nonmagnetic) phase ($V_{PM}(T)$) normalized with respect to the paramagnetic unit cell volume ($V_{PM}(T)$). It can be expressed (as shown by equation-8 in main text) as $\lambda_{ms}^V(T) = \frac{V_{AFM}(T) - V_{PM}(T)}{V_{PM}(T)}$. The $\lambda_{ms}^V$ estimated from powder x-ray diffraction measurements (PXRD) as a function of temperature is shown in Figure S16.

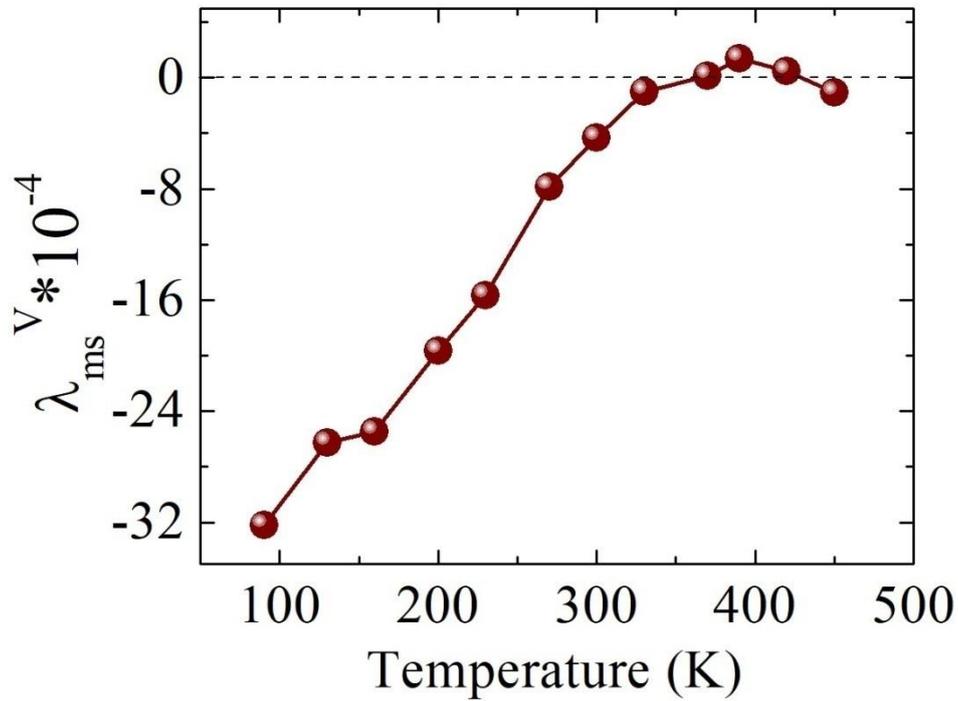

Figure S16. The spontaneous volume magnetostriction ($\lambda_{ms}^V$) of 15R-BaMnO₃ as a function of temperature (estimated from PXRD).

*Estimation of Grüneisen parameter*

We have calculated $\frac{\Delta\omega}{\omega} = \frac{\omega_T - \omega_{80K}}{\omega_{80K}}$ and $\frac{\Delta V}{V} = \frac{V_T - V_{80K}}{V_{80K}}$ from temperature-dependent Raman and XRD results, respectively (Figures S17 and S18). Considering the relation $\frac{\Delta\omega}{\omega}(T) = -\gamma \frac{\Delta V}{V}(T)$, we have estimated the Grüneisen parameter ($\gamma$) for all the modes which show linear relation between $\frac{\Delta\omega}{\omega}(T)$ and $\frac{\Delta V}{V}(T)$ in the temperature range of 80-330 K as shown in Figure S19 and the value is found to be in between 1 and 2 (displayed in Figure S19).

*Magnetostriction from Raman measurements*

The magnetostriction is defined as the change in the physical dimension caused by magnetization from the demagnetized state to magnetic saturation. Magnetostriction constant is represented by $\frac{\Delta l}{l}$, the change in the length by original length at zero field. Here, we have measured temperature-dependent magnetization under different magnetic fields in the range of 1-7T. The bifurcation between ZFC and FC curves is observed to decrease with applied magnetic field and then completely merges at 7T (Figure S20) which means that the antiferromagnetic state switches to ferromagnetic state at higher magnetic fields. We have collected Raman spectra in this range (1-9T) and estimated the change in phonon frequency due to the magnetic field at a fixed temperature (T=4 K) using $\frac{\Delta \omega}{\omega} = \frac{\omega_H - \omega_{0T}}{\omega_{0T}}$. The change in the phonon frequency associates with a change in volume due to applied magnetic field through the respective Grüneisen parameter ($\gamma$): $\frac{\Delta \omega}{\omega}(H) = -\gamma \frac{\Delta V}{V}(H)$. Under the assumption of $\gamma \sim 1$, $\frac{\Delta V}{V}(H) = -\frac{\Delta \omega}{\omega}(H)$, therefore, the volume magnetostriction is $\lambda_V = \frac{\Delta V}{V}(H) \approx -\frac{\Delta \omega}{\omega}(H)$. The value of $\lambda_V$ for the different phonons is found to be in the range of 3 x$10^{-4}$ to 14 x $10^{-4}$ (Figure S21). The values of $\lambda_V$ found for 15R-BaMnO$_3$ obtained from our results are comparable to several reported values for polycrystalline materials as listed in Table S5 [7-14]. Further, the values of magnetostriction obtained from x-ray and Raman measurements are comparable.

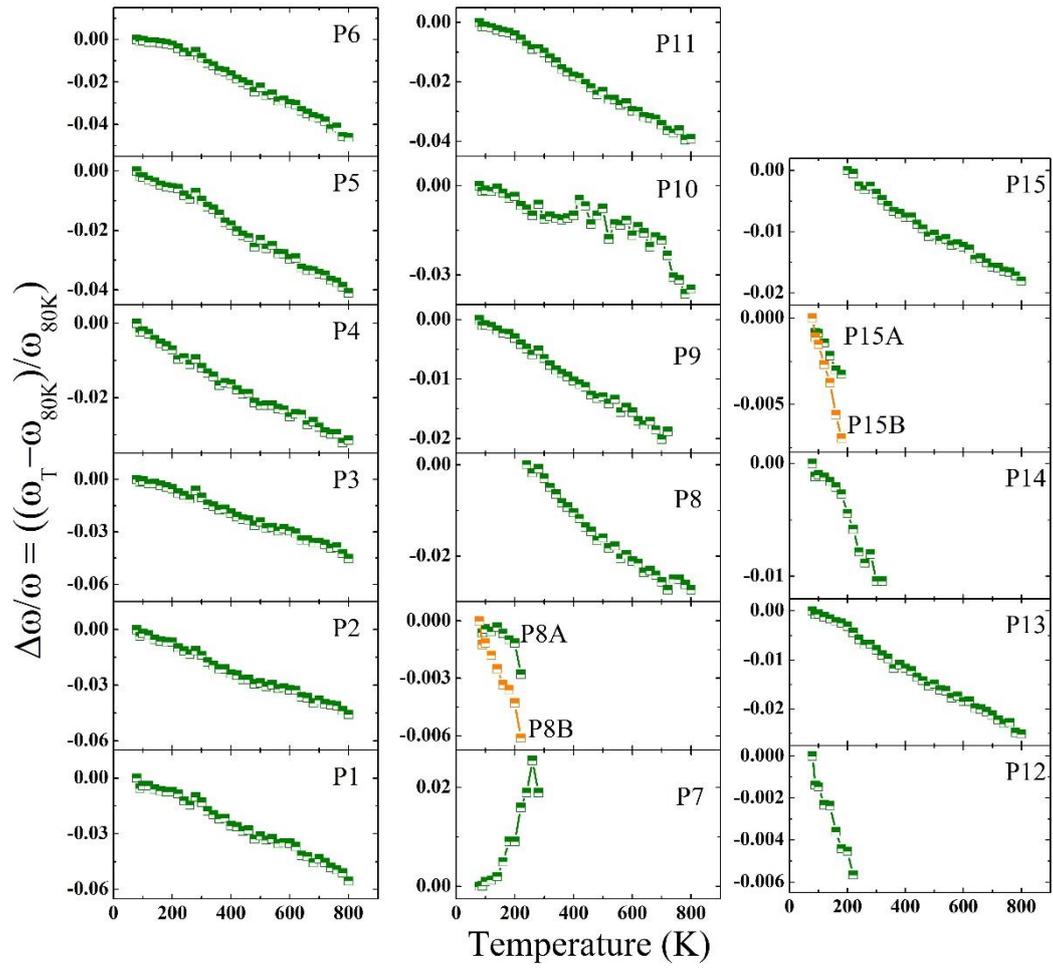

Figure S17. Plot of $\frac{\Delta\omega}{\omega}$ *vs* temperature.

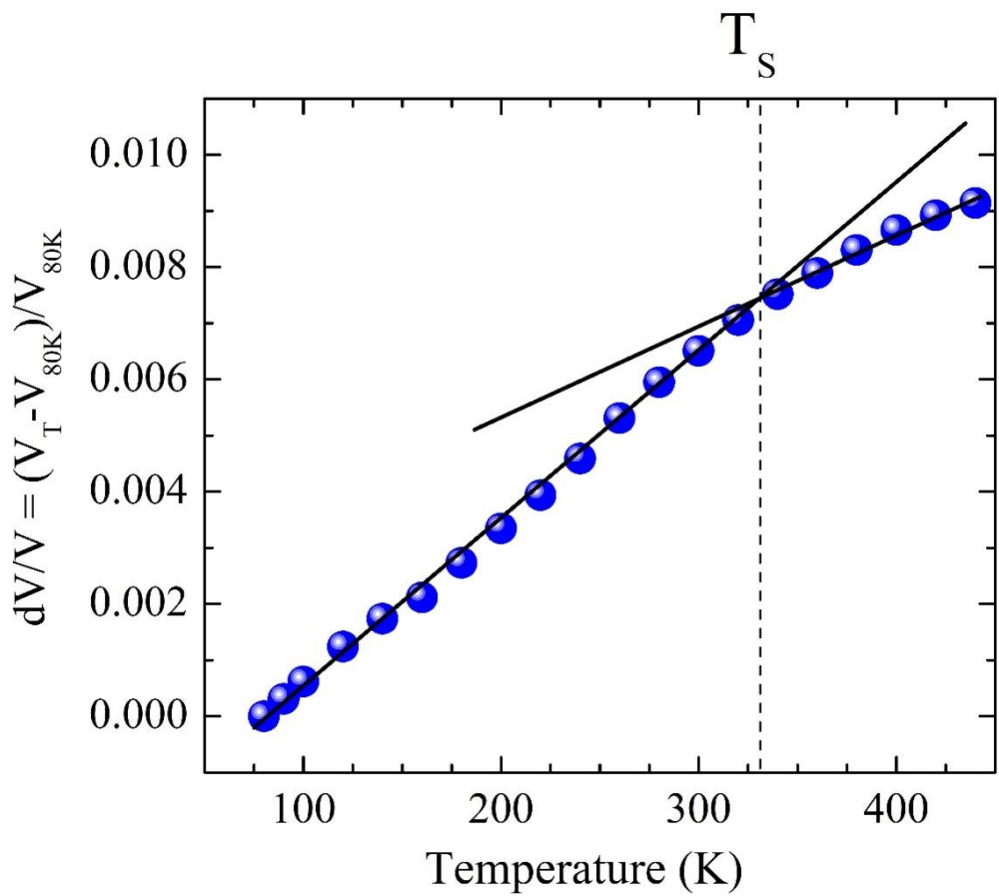

Figure S18. Plot of $\frac{\Delta V}{V}$ vs Temperature. Solid lines are guide to eye showing the deviation at $T_N$.

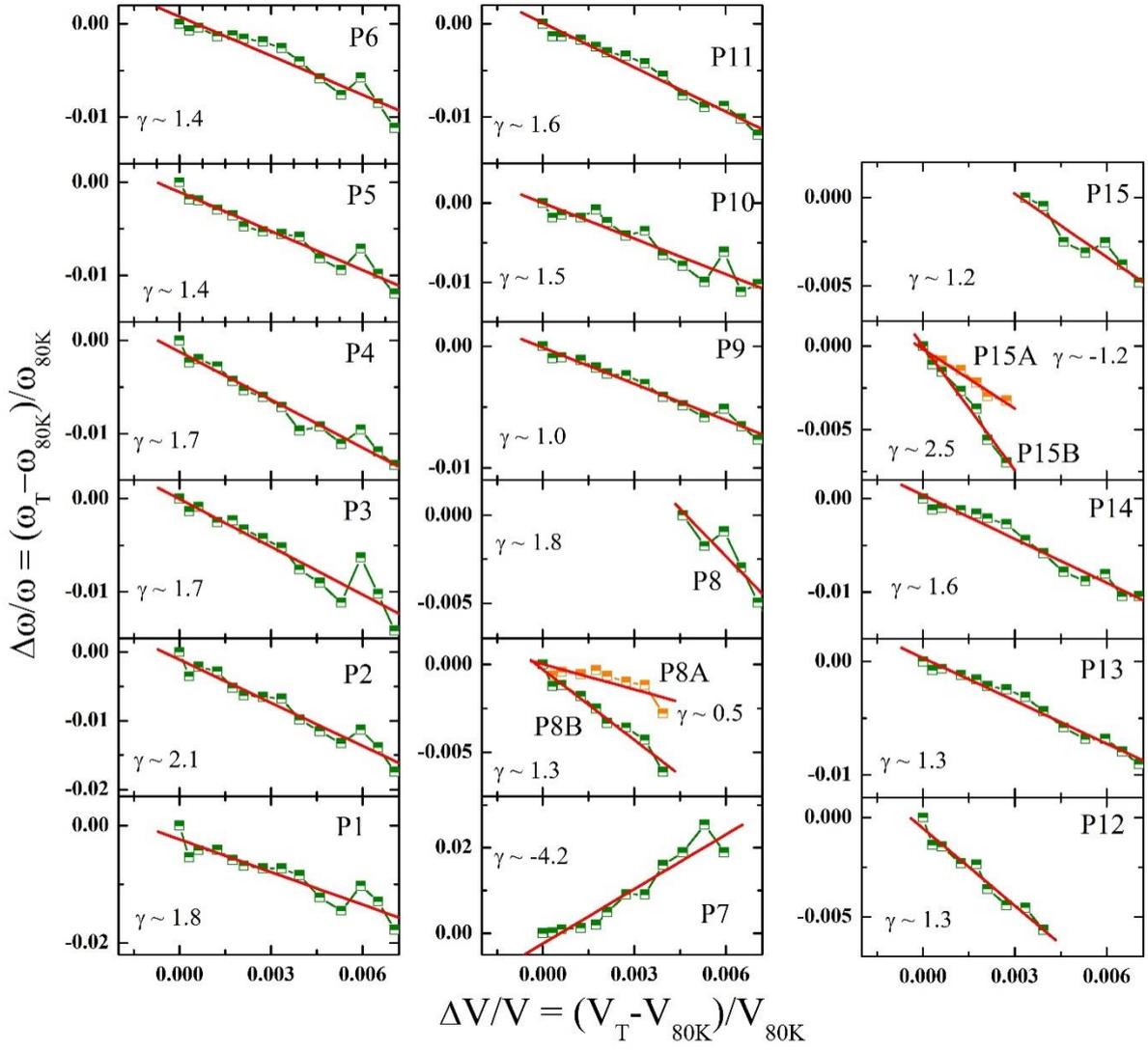

Figure S19. Plot of $\frac{\Delta\omega}{\omega}$ vs $\frac{\Delta V}{V}$. Slope of this plot gives the estimation of Grüneisen parameter ($\gamma$).

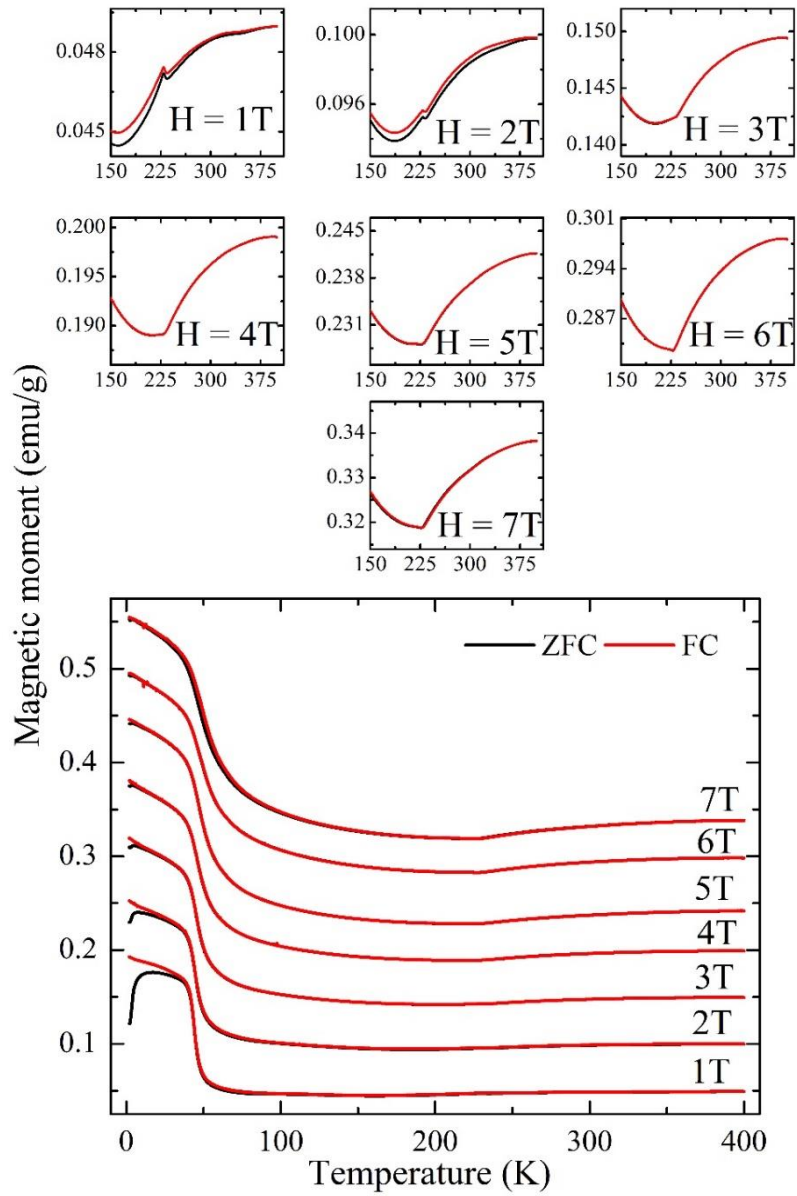

Figure S20. Temperature-dependent magnetization (ZFC and FC) measures under different applied dc magnetic fields.

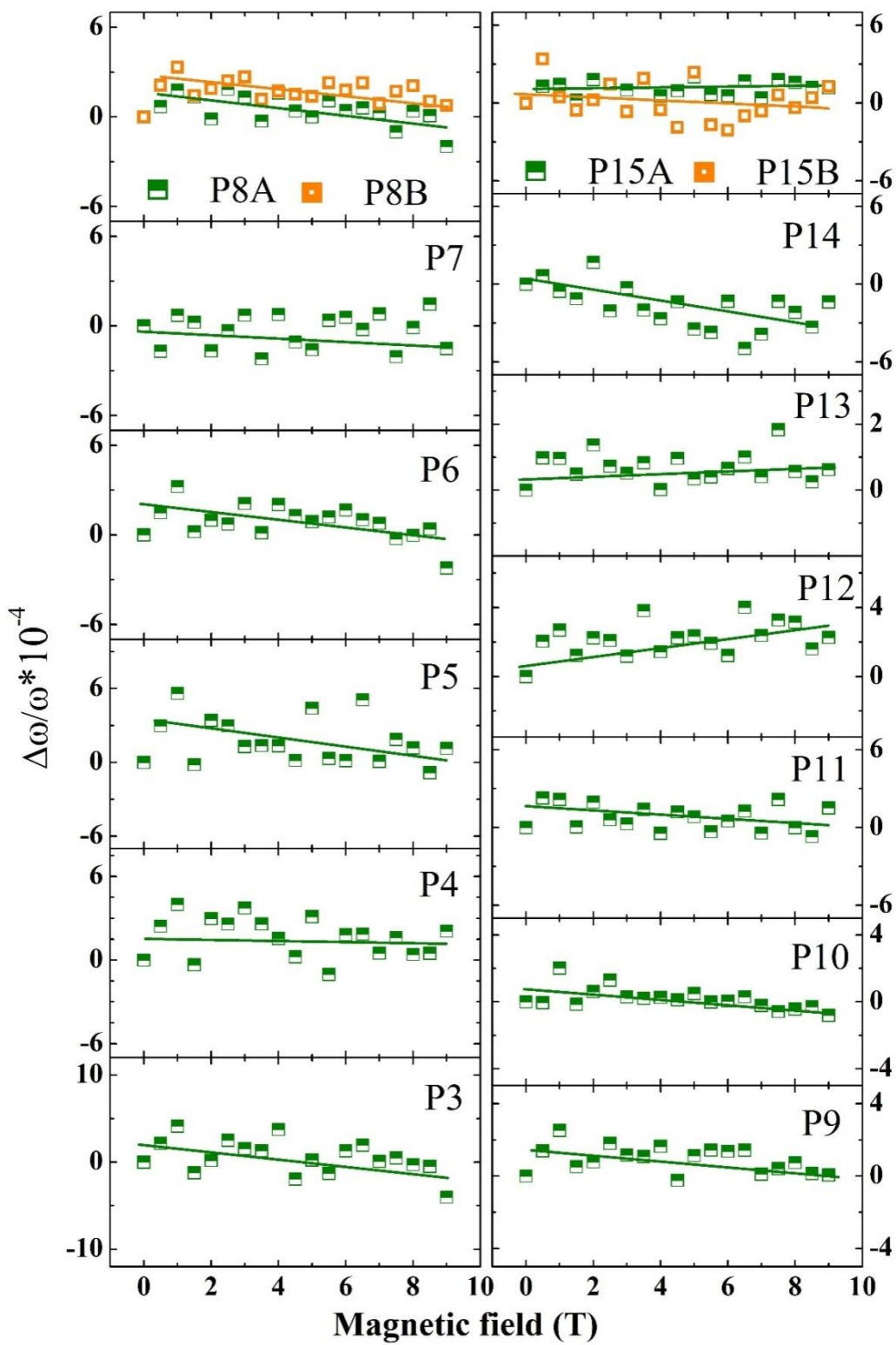

Figure S21. The plot of $\frac{\Delta\omega}{\omega} = \frac{\omega_H - \omega_{0T}}{\omega_{0T}}$ as a function of magnetic field. Solid lines are guide to eye.

Table S5. Values for magnetostriction for a few compounds available in literature for the comparison.

| System | Magnetostriction ($*10^{-6}$) |
|---|---|
| $CoMg_xFe_{2-x}O_4$ | 77 to 221 [Ref. 9] |
| Ce | 65 [Ref. 10] |
| Eu | 70 [Ref. 10] |
| $CoFe_2O_4$ | -110 [Ref. 16] |
| $MnFe_2O_4$ | -5 [Ref. 16] |
| $MgFe_2O_4$ | -6 [Ref. 16] |
| $NiFe_2O_4$ | -24 (along 111) -51 (along 100) (at 4.2 K) [Ref. 14] |
| $Zn_{1-x}Cu_xCr_2Se_4$ | 400 to 2500 (at 100 K) [Ref. 15] |